\pdfoutput=1
\RequirePackage{ifpdf}
\ifpdf % We are running pdfTeX in pdf mode
\documentclass[pdftex]{sigma}
\else
\documentclass{sigma}
\fi

\usepackage{bbm}

\newcommand*{\1}{\mathbbm{1}}
\newcommand*{\dd}{\textrm{d}}

\newcommand*{\mt}{\textrm}
\newcommand*{\mc}{\mathcal}
\newcommand*{\mf}{\mathfrak}
\newcommand*{\mb}{\mathbb}
\newcommand*{\tr}{\operatorname{tr}}

\numberwithin{equation}{section}

\begin{document}

\allowdisplaybreaks

\newcommand{\arXivNumber}{1605.03942}

\renewcommand{\PaperNumber}{006}

\FirstPageHeading

\ShortArticleName{Spacetime-Free Quantum Theory and Ef\/fective Spacetime Structure}

\ArticleName{Spacetime-Free Approach to Quantum Theory\\ and Ef\/fective Spacetime Structure}

\Author{Matti RAASAKKA}

\AuthorNameForHeading{M.~Raasakka}

\Address{N\"ayttelij\"ankatu 25, 33720 Tampere, Finland}
\Email{\href{mailto:mattiraa@gmail.com}{mattiraa@gmail.com}}
\URLaddress{\url{https://sites.google.com/site/mattiraa/}}

\ArticleDates{Received May 13, 2016, in f\/inal form January 17, 2017; Published online January 24, 2017}

\Abstract{Motivated \looseness=1 by hints of the ef\/fective emergent nature of spacetime structure, we formulate a spacetime-free algebraic framework for quantum theory, in which no a priori background geometric structure is required. Such a framework is necessary in order to study the emergence of ef\/fective spacetime structure in a consistent manner, without assuming a background geometry from the outset. Instead, the background geometry is conjectured to arise as an ef\/fective structure of the algebraic and dynamical relations between observables that are imposed by the background statistics of the system. Namely, we suggest that quantum reference states on an \emph{extended} observable algebra, the free algebra generated by the observables, may give rise to ef\/fective spacetime structures. Accordingly, perturbations of the reference state lead to perturbations of the induced ef\/fective spacetime geometry. We initiate the study of these perturbations, and their relation to gravitational phenomena.}


\Classification{81T05; 83C45; 81P10; 81R15; 46L09; 46L53}

%\tableofcontents

\medskip

\rightline{\emph{Dedicated to the memory of Ukki (1920--2015).}}

\section{Introduction}
\looseness=1 The human species has evolved to thrive in the low-energy regime of the Universe, where the environment can be very ef\/fectively described by classical geometry. Our brains are thus hard-wired for geometrical thinking, which is strongly ref\/lected in the historical development of mathematics and physics. Indeed, our best understanding of spacetime today, Einstein's general theory of relativity, is entirely based on a geometric description of its structure. However, for more than a century now, indications have kept emerging that geometry may not be a suitable framework for describing the behavior of spacetime at very high energies~-- quantum ef\/fects seem to undermine the geometric description of spacetime. This poses a deep challenge for theoretical physics, not least because we may no longer be able to rely on our innate geometrical intuition.

Quantum theory imposes fundamental limitations on the accuracy of spacetime measurements. For example, we cannot approximate with arbitrary precision free test point particles, with which spacetime structure may be measured according to general relativity~\cite{Sudarsky07}. On the other hand, considerations of quantum mechanical clocks reveal fundamental limitations to measurements of duration and distance \cite{Gambini07, GiovannettiLloydMaccone04,Lloyd12,Peres80, Salecker58, Tralle05}. Similarly, quantum f\/ield theory and gravitation together prevent the exact localization of events \cite{DoplicherFredenhagenRoberts95}. Therefore, the physical meaning of a spacetime point, and accordingly that of a spacetime manifold, is seriously undermined~\cite{ButterfieldIsham01}. Notably, such considerations also seem to imply that a quantum theory fully incorporating gravity should not follow by directly quantizing the inherently classical manifold structure~\cite{Sudarsky07}.\footnote{Notice, however, that quantizing perturbations of spacetime geometry can still make sense as an ef\/fective f\/ield theory with a limited range of validity, in the same way that, e.g., quantizing density perturbations (i.e., phonons) is sensible for some condensed matter systems, even if spacetime structure were not fundamentally quantum.}

But if geometry cannot be utilized for the fundamental description of spacetime, what can we substitute in its place? Many suggestions have been made since the problem of quantum gravity was f\/irst considered in the 1930s~\cite{Bronstein36}~-- too many to list here (however, see, e.g.,~\cite{Oriti09}). Here we wish to explore a dif\/ferent approach, in which we try to avoid substituting any additional structure.

Let us begin by considering what we fundamentally mean by the notion of spacetime structure. Physics is always tied to what can be observed, and therefore we wish to take the operational approach to the question: How do we measure spacetime structure? In fact, we never directly observe spacetime, but we deduce its structure indirectly by studying the propagation of matter and radiation. Accordingly, we are led to suspect that operational spacetime structure may already be encoded into the structure of the theory that describes the latter, namely, quantum f\/ield theory (QFT). Several hints and suggestions to this ef\/fect have already appeared in the past literature, on which our present work is partly based (see, in particular, \cite{Aguilar12,Bannier94,BertozziniContiLewkeeratiyutkul10,Bianchi12,CorichiRyanSudarsky02,Jacobson15,Keyl98,SummersWhite03}).

We should note that the argument above could be said of any other f\/ield just as well, as the description of physical measurements can always be cast in terms of interactions between dif\/ferent physical systems. However, there are at least two important aspects distinguishing the gravitational f\/ield: (1) We have been able to give a succesful quantum description to the other f\/ields~-- operational information about them is already encoded into the structure of QFT or, more specif\/ically, the Standard Model. (2) The gravitational f\/ield af\/fects the behavior of matter universally, by af\/fecting spacetime structure itself, thus being completely independent of the probe we use for measuring it.\footnote{As is well-appreciated, the universality of gravity is exactly what enables us to describe it in terms of spacetime structure in the f\/irst place, and not as just another f\/ield in spacetime~\cite{Wald84}.} In any case, the above observation simply aims to make plausible the expectation that spacetime geometry could be encoded into the structure of QFT, thus potentially removing the need for an explicit quantization of gravity. It should not be taken as an argument against other possibilities. Also, we do not mean to imply that the other f\/ields could not be understood in a more operational way.

Let us recall some results supporting the idea that gravity may emerge as an ef\/fective phenomenon from a more fundamental quantum description. The f\/irst glimpses of the emergent nature of gravitational phenomena go all the way back to the following realization by Sakharov in 1968 \cite{Sakharov68,Visser02}: Consider QFT on a spacetime with an arbitrary but f\/ixed metric structure coupled to the f\/ield via the covariant derivative (plus possibly a non-minimal term). Then, generically, the Einstein--Hilbert action of general relativity along with a cosmological constant and some higher order corrections can be seen to arise from the one-loop contribution to the ef\/fective action. The derivation is far from unproblematic, since the ef\/fective couplings diverge in the absence of a UV regulator, and the cosmological term comes out far too large even with a~re\-gu\-lator in place. Nevertheless, it does strongly suggest the possibility of an ef\/fective gravitational dynamics arising from quantum corrections. We may hope that by formulating a~better-behaved framework for quantum theory, perhaps with some natural cut-of\/f to the degrees of freedom, gravity may emerge from ef\/fective quantum dynamics.

Another closely related approach to emergent gravity originates from the remarkable thermodynamical properties of black holes that were discovered in the 1970's by Bekenstein \cite{Bekenstein73, Bekenstein74, Bekenstein75} and Hawking \cite{Hawking76}. QFT calculations on a black hole background revealed that the event horizon emits thermal radiation, which gives a temperature and an associated entropy to the black hole. Since then, it has been further understood that such thermodynamical properties can be asso\-cia\-ted not only to black holes but to generic causal horizons in spacetime. As was f\/irst discovered by Unruh~\cite{Unruh76}, even a uniformly accelerating (Rindler) observer in Minkowski spacetime expe\-rien\-ces the thermal behavior of her apparent causal horizon. The origin of these thermodynamical properties of causal horizons continues to be under vigorous debate. However, some evidence indicates that the horizon entropy may be most naturally understood as the entanglement entropy of the quantum f\/ields arising from the correlations between matter degrees of freedom separated by the causal horizon~-- especially if one expects the full gravitational action to be induced by the matter f\/ields \cite{Bombelli86,Cooperman13,Jacobson12,Solodukhin11,Srednicki93}. It was discovered by Jacobson that the Einstein f\/ield equation of general relativity can be obtained as an equation of state for the thermal properties of local Rindler horizons~\cite{Jacobson95}. This relation appears to be a generic feature of gravitational theo\-ries~\cite{Chirco09,Padmanabhan09}. More recently, Jacobson also showed that the semiclassical Einstein equation can be derived in an elegant way from the hypothesis that the QFT vacuum state locally maximizes the von Neumann entropy~\cite{Jacobson15}. Also in this case it is most natural for the Einstein equation to arise solely from the entanglement of matter degrees of freedom. Therefore, there seems to be no need for incorporating gravitons in the theory~-- in fact, the introduction of gravitons into QFT is actively discouraged, as it would lead to a double-counting of energy in Jacobson's derivation~\cite{Jacobson15}.

Let us emphasize that our brief review above of mechanisms for the emergence of gravity from QFT is far from exhaustive. (See, e.g., \cite{Dreyer06,Lloyd12,Padmanabhan12,Sindoni11,Swingle14} and references therein for some of the other approaches.) These mechanisms for the emergence of gravity from quantum f\/ield theory may not all be mutually exclusive, although the relations between them are not well understood at the moment, as far as we know. Nevertheless, we may argue that none of these mechanisms, or indeed any mechanism based on QFT, can present a logically coherent explanation for the emergence of gravity as long as the fundamental quantum theory is built on a background spacetime: According to general relativity, gravity is an inherent property of spacetime~\cite{Wald84}. Therefore, in order to provide a consistent description of the emergence of gravity, spacetime itself must be emergent, and not appear as a fundamental ingredient in the construction of quantum theory.

Our work aims to provide a concrete mathematical framework, a spacetime-free formulation of quantum theory, in which questions of spacetime emergence can be directly and explicitly addressed.\footnote{Perhaps it would be more accurate to call our approach `background-geometry-free' instead of `spacetime-free', since there exist QFT models for quantum gravity (e.g., group f\/ield theories~\cite{Oriti11}), which are formulated on auxiliary spaces not directly related to spacetime, whereas we want to fully remove the geometric background manifold, on which QFT is formulated. However, we opt for the `spacetime-free' terminology for the sake of compactness.} In order to have a chance of success, the framework we wish to develop should satisfy at least the following three requirements:
\begin{enumerate}\itemsep=0pt
	\item Spacetime structure should not enter the theory as a fundamental ingredient. Instead, we should be able to recover spacetime as an ef\/fective description of the dynamical organization of the degrees of freedom in some regime of the theory.
	\item Despite the lack of background spacetime, the theory should have a clear operational interpretation in terms of (idealized) experimental arrangements and observations.
	\item When \looseness=1 an ef\/fective geometric background \emph{can} be recovered, the theory should reproduce in the appropriate regimes our current models, general relativity and quantum f\/ield theo\-ry.
\end{enumerate}

The development of the spacetime-free framework for quantum physics in Section \ref{sec:framework} is guided by these three requirements. We will also assume the general validity of abstract algebraic quantum theory \cite{Araki99,Haag96, Strocchi08}, which we will consider \emph{for all practical purposes} as a theory of knowledge, although the framework itself is independent of the interpretation of the quantum state.\footnote{However, see \cite{Hoehn14b,Hoehn15} for a reconstruction of qubit quantum theory from operational considerations that are rather reminicent of our development of the spacetime-free framework in Section \ref{sec:framework}.}

The following diagram depicts the traditional logic of constructing a quantum f\/ield theoretic model, according to which we f\/irst postulate a background spacetime, and then build the QFT model on top of it:
\begin{gather*}
	\mt{spacetime}\ \stackrel{\mt{locality}}{\Longrightarrow}\ \mt{dynamics}\ \stackrel{\mt{KMS cond.}}{\Longrightarrow}\ \mt{equilibrium/vacuum state}
\end{gather*}
The arrows in this diagram should be understood as one-to-many correspondencies. For example, the notion of locality provided by the background geometry does not completely f\/ix the dynamics, but strongly restricts its choice for the physically relevant local QFT models, such as the Standard Model. Obviously, there exist many local QFT models corresponding to the same background geometry. Similarly, the same dynamics give rise to several dif\/ferent equilibrium states parametrized by temperature.\footnote{Let us neglect for the moment the fact that not all spacetimes allow for equilibrium states. We will later def\/ine the concept of a reference state that need not be an equilibrium/vacuum state.} As an example, the locally covariant QFT framework~\cite{BrunettiFredenhagenVerch03,FewsterVerch11}, which is arguably the most general formulation of ordinary QFT today, follows this logic in def\/ining QFT models. Following ideas in earlier works~\cite{Aguilar12,Bannier94,BertozziniContiLewkeeratiyutkul10,Bianchi12,CorichiRyanSudarsky02,Keyl98,SummersWhite03}, we suggest trying to invert the arrows leading from the spacetime to the equilibrium state in order to recover spacetime geometry from the equilibrium state.\footnote{During the revision of this manuscript, we became aware of the interesting paper~\cite{Salehi97} by Salehi, which also explores the idea of state dependent dynamics for quantum systems.} The purpose of this reversal of logic is that a state of the system can be determined and represented algebraically without referring to any spacetime structure. Hence, our diagram would look more like this:
\begin{gather*}
	\mt{spacetime}\ \stackrel{\mt{???}}{\Longleftarrow}\ \mt{dynamics}\ \stackrel{\mt{T-T theory}}{\Longleftarrow}\ \mt{equilibrium/vacuum state}
\end{gather*}
The second arrow from the dynamics to the equilibrium state is already known (in some cases) to be inverted by the Tomita--Takesaki modular theory, as reviewed in Section~\ref{subsec:modular}. The f\/irst arrow leading from the dynamics to the spacetime is more enigmatic, although some ideas and hints on how to achieve the inversion have appeared in the literature (see, e.g., \cite{Bannier94,BertozziniContiLewkeeratiyutkul10,Bianchi12,CaoCarrollMichalakis16,Keyl98,SummersWhite03}). In Section~\ref{sec:spacetime} we will develop certain physically motivated ideas and methods for studying the ef\/fective spacetime structure induced by the dynamics.

\looseness=1 The rest of this paper is organized as follows. The Sections~\ref{subsec:free} and~\ref{subsec:refstate} form the core content of the paper, in which we give a mathematically precise def\/inition for the spacetime-free quantum theory. The ability of the formulation to describe general quantum systems, even in the absence of spacetime structure, relies on the universality property of the free product of algebras, as explained in Section~\ref{subsec:free}. In the rest of the Section~\ref{sec:framework} we review several canonical operator algebraic structures appearing in the spacetime-free framework that should play a key role in the extraction of dynamics and ef\/fective spacetime structure from the organization of the quantum statistics of observations. In Section~\ref{sec:spacetime} we further of\/fer some ideas on how to actually recover information on the ef\/fective spacetime geometry. This section mainly serves the purpose of making it at least plausible to the reader that such a recovery is possible, even though we do not yet have solid results to of\/fer on this aspect of the approach. Section~\ref{sec:conclusions} provides a summary of the results, and points out some of the challenges and future prospects for the approach. This paper also contains several appendices that of\/fer further details to the presentation.

\section{Spacetime-free framework for quantum physics}\label{sec:framework}
In relativistic quantum f\/ield theory, the causal structure of spacetime is encoded into the algebraic structure of the observable algebra~-- in particular, commutators of spacelike separated observables vanish~\cite{Haag96}. On the other hand, from general relativity we know that the matter distribution of a system has an inf\/luence on the causal structure through gravity~\cite{Wald84}. In quantum f\/ield theory the matter distribution is described by the quantum state. Therefore, in order to introduce gravitational ef\/fects into quantum f\/ield theory, we must be able to modify the for\-ma\-lism so that the quantum state may inf\/luence the algebraic relations of the observables. To this end, we f\/irst introduce the free observable algebra, which will allow us to give an operational def\/inition of a quantum system in the absence of a background spacetime.

\subsection{Free observable algebra}\label{subsec:free}
To def\/ine an experimental arrangement we introduce the set $\mc{M} := \{M_i \colon i\in I\}$ of measurements that can be performed in the experiment, where $I$ is an arbitrary index set.\footnote{The measurements can also be understood as partial observables of the system under study, as def\/ined in~\cite{Rovelli01}.} To any measurement $M \in\mc{M}$ we associate a spectrum $\mt{Spec}(M)$, which is the topological (compact Hausdorf\/f) space of possible values that the measurement can take.\footnote{Typically, $\mt{Spec}(M) \subset \mb{R}$ of course, but we need not to make such restriction here. However, we do assume for simplicity that $\mt{Spec}(M)$ is compact for all $M\in\mc{M}$. If $\mt{Spec}(M)$ is not compact, we can restrict to functions that vanish at inf\/inity in the following, and the construction goes through more or less the same way.} We consider the spectrum $\mt{Spec}(M)$ to fully characterize the measurement $M\in\mc{M}$. We may then associate to the measurement a~unital abelian $C^*$-algebra $\mf{M} := C(\mt{Spec}(M),\mb{C})$ that is the algebra of continuous complex-valued functions on its spectrum. Continuous real-valued functions on $\mt{Spec}(M)$ correspond to self-adjoint operators in $\mf{M}$, which represent the observables that are accessible through the measurement $M \in \mc{M}$. By the Gelfand duality, $\mf{M}$ is uniquely determined by $\mt{Spec}(M)$, and vice versa. (See, e.g.,~\cite{Varilly06} for an elementary exposition of the Gelfand duality.) In order to describe projective measurements, we may complete $\mf{M}$ in the weak operator topology to obtain the abelian von Neumann algebra $\mf{W}$, which contains the spectral projections $\{P_{O}\}_{O\subset \mt{Spec}(M)}$, where $O \subset \mt{Spec}(M)$ are open sets. An abelian von Neumann algebra can be identif\/ied as a~(classical) probability space, where the projection $P_{O}$ corresponds to the proposition that the value of the random variable resides in the open set $O\subset \mt{Spec}(M)$. By extension, non-abelian von Neumann algebras are often considered to be non-commutative probability spaces. (The books~\cite{KadisonRingrose,KadisonRingrose2, Takesaki1,Takesaki2,Takesaki} of\/fer extensive references for operator algebra theory, and \cite{Araki99,Haag96,Strocchi08} give excellent accounts of algebraic quantum theory.)

A central mathematical construct for the development of our framework is the free product of algebras\footnote{Throughout this paper, we refer by the term `free product' to what is more precisely called the \emph{reduced} free product, where unit elements of the component algebras are identif\/ied.}, which plays a fundamental role in the non-commutative probability theory~\cite{Speicher11, Voiculescu92}. The free product $\mf{A}_1\star\mf{A}_2$ of two unital $*$-algebras $\mf{A}_i$, $i=1,2$, is linearly generated by f\/inite sequences $x_1x_2\cdots x_n$, where $x_k\in\mf{A}_1$ or $x_k\in\mf{A}_2$ for each $k \in 1,\ldots,n$. The product of two elements $x_1\cdots x_m, y_1\cdots y_n \in \mf{A}_1\star\mf{A}_2$ is given by the concatenation of sequences, i.e.,
\begin{gather*}
	(x_1\cdots x_m) \star (y_1\cdots y_n) = x_1\cdots x_m y_1\cdots y_n .
\end{gather*}
Moreover, the involutive ${}^*$-operation is given by $(x_1\cdots x_m)^* = x_m^*\cdots x_1^*$. Finally, we impose the following equivalence relations on the elements of $\mf{A}_1\star\mf{A}_2$:
\begin{enumerate}\itemsep=0pt
	\item $\1_{\mf{A}_1} \sim \1_{\mf{A}_2} \sim \1_{\mf{A}_1\star\mf{A}_2}$, the unit element (i.e., the empty sequence) in $\mf{A}_1\star\mf{A}_2$, where $\1_{\mf{A}_i}$ denotes the unit element in $\mf{A}_i$, and $\sim$ implies equivalence.
	\item If $x_k,x_{k+1} \in \mf{A}_i$ for the same $i=1,2$, we set $x_1\cdots x_k x_{k+1} \cdots x_n \sim x_1\cdots (x_k \cdot x_{k+1}) \cdots x_n$, where in the latter we denote by $(x_k \cdot x_{k+1})\in\mf{A}_i$ the product of $x_k$ and $x_{k+1}$ in $\mf{A}_i$.
\end{enumerate}
The free product of $*$-algebras is an associative and commutative operation in the category of $*$-algebras. The free product $*$-algebra $\mf{A}_1\star\mf{A}_2$ is always non-abelian and inf\/inite-dimensional unless one of the factors is trivial (i.e., isomorphic to~$\mb{C}$). Moreover, it has the following important \emph{universality property}: $\mf{A}_1\star\mf{A}_2$ is the unique unital $*$-algebra, from which there exists a unital $*$-homomorphism to any other $*$-algebra generated by $\mf{A}_1$ and $\mf{A}_2$ as unital $*$-subalgebras~\cite{Voiculescu92}. In this precise sense, the free product does not impose any relations between the elements of its factors except for the identif\/ication of units. Accordingly, it is the canonical structure to start with, when we wish to impose arbitrary relations between unital subalgebras.

Let us then def\/ine the \emph{free observable algebra} $\mf{F}$ associated with an experimental arrangement. It is the unital $*$-algebra given by the free product $\mf{F} := \star_{i\in I} \mf{W}_i$ of the abelian von Neumann algebras $\mf{W}_i$ associated with the measurements $M_i \in\mc{M}$ available in the experimental arrangement. In particular, we interpret an element of the form
$P_{O_1}^{(i_1)} P_{O_2}^{(i_2)}\cdots P_{O_n}^{(i_n)} \in \mf{F}$,
i.e., a sequence of spectral projections $P_{O_k}^{(i_k)} \in \mf{W}_{i_k}$, $O_k\subset \mt{Spec}(M_{i_k})$, to represent a sequence of measurement outcomes associated with the measurements. Importantly, the set of f\/inite sequences of spectral projections linearly generate the free observable algebra, since each of the factors $\mf{W}_i$ is (the norm completion of the algebra) linearly generated by its spectral projections. The unit element $\1_\mf{F}\in\mf{F}$ represents the case of no measurements.

\looseness=-1 Let us emphasize that the ordering of spectral projections refers to the order, in which the mea\-su\-rement results are recorded by the experimenter~-- not (necessarily) the temporal order of the measurement events. Thus, no global time f\/low is needed for the def\/inition of this ordering, but only the assumption that the experimenter can perform sequences of measurements. The ordering of the measurements is relevant for quantum systems due to quantum uncertainty relations: Recording the value of one observable quantity may inf\/luence the results of other measurements.

It may be helpful to contrast the free product with the tensor product to better understand the physical signif\/icance of the free observable algebra. The usual way to combine the observable algebras of individual quantum systems to form the observable algebra of the composite system is the tensor product. However, the tensor product already requires certain assumptions on the relations between the component systems (e.g., that their observable algebras commute mutually), which are equivalent to the operational independence of the systems~\cite{Summers08}. Indeed, usually the quantum systems that are put together by tensor product are considered spacelike separated and/or causally independent. The free product, on the other hand, does not impose any such relations a priori. By the universality property, we may recover any possible algebraic relations between the individual systems via unital $*$-homomorphisms of the free product of their observable algebras. The tensor product structure corresponding to operational independence of the systems is but only one of the vast array of possible relations between the component algebras that can be obtained in this way. Other possibilities include the many ways in which the systems may not be independent or mutually commuting, and thus one can see how these homomorphisms may naturally introduce temporal/causal structure on the composite system. Above, we have restricted to consider the situation in which the individual systems correspond to abelian observable algebras generated by single observables, because we wish not to assume anything about the spatiotemporal relations of the observables a priori, whereas non-abelian algebras may already contain information on the temporal structure or the dynamics of the system.\footnote{It is, of course, possible to generalize the construction of the free observable algebra also to the case where the individual algebras are non-abelian, because the free product can be def\/ined for any family of $*$-algebras.} The tensor product of abelian algebras is again an abelian algebra, and therefore would not give us any interesting algebraic structure in any case, whereas the free product is always an inf\/inite-dimensional non-abelian algebra, if the component algebras are non-trivial (i.e., not isomorphic to~$\mb{C})$.

\subsection{Reference states and the physical observable algebra}\label{subsec:refstate}
In addition to the set of possible measurements, an experiment is always characterized by the statistical background to the measurements, by which we refer specif\/ically to the probabilities and correlations of measurement outcomes in the absence of any additional external interference with the measured system. In ordinary QFT in Minkowski spacetime, for example, the statistical background for the usual particle scattering experiments carried out in a vacuum environment is represented by the vacuum state. Experimentally, the statistical background for an experiment is obtained via the calibration of the measurement instruments prior to any actual measurements. In order to obtain good statistics for the background noise, one must repeat the same calibration experiment a large number of times with a collection of systems prepared in exactly the same way~-- the statistical ensemble. In any realistic situation the experimental background statistics are, of course, always limited in accuracy due to the f\/inite ensemble size. However, in most cases we may believe that the background statistics converges to a limit as the cardinality of the ensemble is increased. If the calibration experiments are performed on the same system but at separate intervals of external time as experienced by the experimenter, while letting the system to `relax' between the experiments, the statistical background must be in equilibrium with respect to the external time f\/low. In a controlled measurement scheme, the measurement results are compared with the statistical background in order to dif\/ferentiate the ef\/fects of external inf\/luence on the system. It is a characteristic property of quantum systems that the statistical background can never be completely trivial, since quantum ef\/fects produce f\/luctuations and correlations in measurement outcomes even in the pure vacuum.

As usual in algebraic quantum theory, we will use states to represent information about the experimental system. (See, e.g., \cite{Araki99,Haag96, Strocchi08} for comprehensive accounts of the algebraic formulation of quantum theory.) A \emph{state} $\omega$ on the free observable algebra $\mf{F}$ is a linear functional $\omega\colon \mf{F} \rightarrow \mb{C}$, which is\footnote{Note that not all of these requirements are completely independent: For example, the positivity of a state is enough to guarantee self-adjointness and continuity in the case of $C^*$-algebras.}
\begin{enumerate}\itemsep=0pt
	\item[1)] positive: $\omega(a^*a) \geq 0$ for all $a\in\mf{F}$,
	\item[2)] self-adjoint: $\omega(a^*) = \overline{\omega(a)}$ for all $a\in\mf{F}$,
	\item[3)] normalized: $\omega(\1_\mf{F}) = 1$, and
	\item[4)] independently continuous in each of the elements $x_1,\ldots,x_n$ for any sequence $x_1\cdots x_n\in\mf{F}$.
\end{enumerate}
Let us denote the space of states on $\mf{F}$ by $\mc{S}(\mf{F})$. Since the free observable algebra is generated by f\/inite sequences of spectral projections, the values on these elements determine the state. For any state $\omega \in \mc{S}(\mf{F})$ we interpret
\begin{gather}\label{eq:probs}
	\omega\Big( \big(P_{O_1}^{(i_1)} P_{O_2}^{(i_2)}\cdots P_{O_n}^{(i_n)}\big)^* \big(P_{O_1}^{(i_1)} P_{O_2}^{(i_2)}\cdots P_{O_n}^{(i_n)}\big) \Big) \in [0,1]
\end{gather}
as the probability for the sequence of measurement events corresponding to the sequence of spectral projections $P_{O_1}^{(i_1)} P_{O_2}^{(i_2)}\cdots P_{O_n}^{(i_n)}$. This probability interpretation is standard in quantum theory (see Appendix~\ref{app:histories}). However, here we propose to generalize equation~(\ref{eq:probs}) to the case in hand, where no global time parameter or evolution is available a priori. In fact, the generalization asks to distinguish between the time-ordering of events as recorded by the experimenter, which is represented by the ordering of the spectral projections, and the time-evolution of the experimental system. From the point of view of general relativity, it is natural and expected that such a distinction should be made due to the lack of a global time-ordering of events that would tie together the time of the experimenter and that of the system.\footnote{The distinction seems also natural from the point of view of the Bayesian interpretation of the quantum state (see, e.g.,~\cite{Brukner03,Fuchs14}), according to which the state represents the subjective knowledge of the experimenter about the quantum system.}

Next we review very brief\/ly the Gelfand--Naimark--Segal (GNS) construction for a state \mbox{$\omega \in \mc{S}(\mf{F})$}. (See, e.g.,~\cite{Schmudgen90} for a more complete exposition.) Let us denote the \emph{null ideal} of~$\omega$ by
\begin{gather*}
	\mc{N}_\omega := \big\{a \in \mf{F} \colon \omega(a^*a) = 0\big\} ,
\end{gather*}
which is a left ideal in $\mf{F}$. Thus, we may consider the quotient
\begin{gather*}
	\mf{F}/\mc{N}_\omega := \big\{a + \mc{N}_\omega \colon a \in \mf{F}\big\}
\end{gather*}
as a linear space. Denote by $|a\rangle_\omega \in \mf{F}/\mc{N}_\omega$ the equivalence class containing the element $a\in\mf{F}$. Then, $\omega$ provides a non-degenerate inner product
\begin{gather*}
	\langle a | b \rangle_\omega := \omega(a^*b)\qquad \forall\, |a\rangle_\omega, |b\rangle_\omega \in \mf{F}/\mc{N}_\omega
\end{gather*}
on $\mf{F}/\mc{N}_\omega$. We may now complete $\mf{F}/\mc{N}_\omega$ in the norm induced by this inner product to obtain the GNS Hilbert space $\mc{H}_\omega := \overline{\mf{F}/\mc{N}_\omega}$. Setting
\begin{gather*}
	\pi_\omega(a) |b\rangle_\omega \stackrel{!}{=} |ab\rangle_\omega \in \mf{F}/\mc{N}_\omega
\end{gather*}
for all $a\in \mf{F}$ and $|b\rangle_\omega \in \mf{F}/\mc{N}_\omega$ gives rise to the GNS representation $\pi_\omega\colon \mf{F} \rightarrow B(\mc{H}_\omega)$ of $\mf{F}$ on the whole of $\mc{H}_\omega$ by continuity. We have by construction that the unit vector $|\1_\mf{F}\rangle_\omega \in \mf{F}/\mc{N}_\omega$ satisf\/ies $\langle \1_\mf{F}| \pi_\omega(a) |\1_\mf{F}\rangle_\omega = \omega(a)$ for all $a\in\mf{F}$. It is also cyclic in $\mc{H}_\omega$ with respect to $\pi_\omega(\mf{F})$, i.e., the subspace
\begin{gather*}
	\big\{\pi_\omega(a)|\1_\mf{F}\rangle_\omega \in \mc{H}_\omega \colon a \in \mf{F} \big\} \subset \mc{H}_\omega
\end{gather*}
is norm dense in $\mc{H}_\omega$. By the von Neumann double-commutant theorem, we may complete $\pi_\omega(\mf{F}) \subset B(\mc{H}_\omega)$ in the weak operator topology on $B(\mc{H}_\omega)$ by taking the double-commutant $\mf{A}_\omega := \pi_\omega(\mf{F})''$, which is therefore a von Neumann algebra.\footnote{The commutant of a subalgebra $\mf{A} \subset B(\mc{H})$ is def\/ined as $\mf{A}' := \{a' \in B(\mc{H}) \colon [a,a'] = 0\ \forall a \in \mf{A}\}$.} The canonical extension $\tilde\omega$ of the state $\omega\in\mc{S}(\mf{F})$ onto $B(\mc{H}_\omega)$ (and thus onto $\mf{A}_\omega$) is obtained as $\tilde\omega(A) := \langle \1_\mf{F}| A |\1_\mf{F}\rangle_\omega$ for all $A \in B(\mc{H}_\omega)$.

Notice that, by the completion procedure, the GNS Hilbert space $\mc{H}_\omega$ also contains any vector of the form $|x_1x_2\cdots\rangle_\omega \in \mc{H}_\omega$, consisting of an inf\/inite number of elements $x_k\in \cup_i\mf{W}_i$ $\forall\, k\in\mb{N}$, associated with the limit of a Cauchy sequence of vectors $(|x_1x_2\cdots x_n\rangle_\omega)_{n\in \mb{N}}$. A similar statement applies to $\mf{A}_\omega$ as a completion of $\pi_\omega(\mf{F}) \subset B(\mc{H}_\omega)$ in the weak operator topology on~$B(\mc{H}_\omega)$.

We will mathematically model the statistical background to an experiment by a special state, the \emph{reference state} $\Omega \in \mc{S}(\mf{F})$. Namely, $\Omega$ encodes the probabilities of measurement outcomes in the absence of any additional external perturbations to the experimental system via equation~\eqref{eq:probs}.\footnote{As an example, for particle scattering experiments in a vacuum environment in Minkowski spacetime, the reference state would be given by the vacuum state (or its restriction to a local subsystem).} The reference state $\Omega$ gives rise to the $*$-representation $\pi_\Omega\colon \mf{F} \rightarrow B(\mc{H}_\Omega)$ of the free observable algebra on the GNS Hilbert space $\mc{H}_\Omega$, and to the von Neumann algebra $\mf{A}_\Omega := \pi_\Omega(\mf{F})'' \subset B(\mc{H}_\Omega)$. We will assume that the restriction of $\Omega$ on each subalgebra $\mf{W}_i \subset \mf{F}$ is faithful, so that $\mf{W}_i$ are represented faithfully in $\mf{A}_\Omega$. We call $\mc{H}_\Omega$ and $\mf{A}_\Omega$ the \emph{physical Hilbert space} and the \emph{physical observable algebra} of the experimental arrangement, respectively. The algebraic structure of the physical observable algebra $\mf{A}_\Omega$ encodes the causal and dynamical properties of the observables under consideration. The free observable algebra together with the reference state form a tuple $(\mf{F},\Omega)$, which we take to fully determine the experimental arrangement.

Crucially for the above, the GNS representation $\pi_\Omega$ as a $*$-homomorphism may impose non-trivial algebraic relations between the factor algebras $\mf{W}_i$ of the free observable algebra $\mf{F} \equiv \star_i \mf{W}_i$, and thus induce non-trivial relations among the observables in $\mf{A}_\Omega$. For example, two obser\-vables or, more generally, two subalgebras of observables $\mf{B}_1,\mf{B}_2 \subset \mf{A}_\Omega$ are \emph{jointly measurable} if they commute in $\mf{A}_\Omega$: $\mf{B}_1 \subset \mf{B}_2'$. Moreover, two subalgebras $\mf{B}_1,\mf{B}_2 \subset \mf{A}_\Omega$ are \emph{operationally independent} if they satisfy the \emph{split property}: There exists a type I von Neumann factor algebra $\mf{C} \subset B(\mc{H}_\Omega)$ such that $\mf{B}_1 \subset \mf{C} \subset \mf{B}_2'$. For operationally independent subalgebras there exist arbitrary normal product states, so the subsystems represented by the two subalgebras can be decorrelated and prepared independently \cite{Summers08}.\footnote{Clearly, operational independence implies joint measurability. For f\/inite-dimensional algebras the two notions coincide.} In this sense, there exists a complete set of operations (represented mathematically by completely positive maps) that af\/fect expectation values of observables only in one of the subalgebras but not the other. Therefore, operational independence is strongly related to causal independence of subsystems~-- indeed, subsystems that are spatially separated by a f\/inite distance are known to be operationally independent in physical QFT models \cite{Summers08}. In the spacetime-free framework, we could even take operational independence as the rigorous def\/inition of causal independence. Notice that operational independence does not imply that there are no statistical correlations between the two subalgebras of observables in an arbitrary state, only that it is possible to remove all correlations.

\subsection{Covariance and symmetries of the experimental arrangement}
Let $\alpha \in \mt{Aut}(\mf{F})$ be a $*$-automorphism of the free observable algebra, and def\/ine $\alpha \triangleright (\mf{F}, \Omega) := (\alpha(\mf{F}), \alpha^*(\Omega))$, where $\alpha^*(\Omega) := \Omega\circ\alpha^{-1}$. We call this action by automorphisms on the tuple a~\emph{covariant transformation}. The covariantly transformed tuple $(\alpha(\mf{F}), \alpha^*(\Omega))$ corresponds to the same experimental arrangement as $(\mf{F}, \Omega)$, because it leads to the same expectation values and thus the same algebraic relations for the physical observable algebra. Therefore, there is a one-to-one correspondence between experimental arrangements and the equivalence classes of tuples by covariant transformations.\footnote{Here, in the def\/inition of the free observable algebra $\mf{F}$, we implicitly include the identif\/ication of the subalgebras $\mf{W}_i$ corresponding to the initial set of physical observables giving rise to $\mf{F} = \star_i \mf{W}_i$.}

For a given reference state $\Omega\in\mc{S}(\mf{F})$, there is a special subgroup
\begin{gather*}
	\mt{Sym}_\Omega(\mf{F}) := \big\{\alpha \in \mt{Aut}(\mf{F}) \colon \alpha^*(\Omega) = \Omega \big\} \subset \mt{Aut}(\mf{F})
\end{gather*}
of automorphisms of $\mf{F}$, the \emph{symmetry group} of the experimental arrangement. Only the free observable algebra transforms under the covariant action by symmetries, $\alpha \triangleright (\mf{F},\Omega) = (\alpha(\mf{F}), \Omega)$. Any symmetry $\alpha\in\mt{Sym}_\Omega(\mf{F})$ can be represented on the GNS Hilbert space $\mc{H}_\Omega$ by a unitary operator $U_\alpha \in B(\mc{H}_\Omega)$, which satisf\/ies $U_\alpha |\1_\mf{F}\rangle_\Omega = |\1_\mf{F}\rangle_\Omega$~\cite{KadisonRingrose,KadisonRingrose2}. On the other hand, we may more generally consider the group of unitaries $\mc{U}_\mf{F} := \{U \in B(\mc{H}_\Omega) \colon U\mf{A}_\Omega U^* = \mf{A}_\Omega \}$ f\/ixing the physical ob\-ser\-vab\-le algebra $\mf{A}_\Omega$, and its subgroup $\mc{U}_\mf{F}^\Omega := \{ U\in \mc{U}_\mf{F} \colon U|\1_\mf{F}\rangle_\Omega = |\1_\mf{F}\rangle_\Omega\}$ that leaves~$\tilde\Omega$ invariant, which is the group of \emph{physical symmetries} of the system. $\mt{Sym}_\Omega(\mf{F})$ is isomorphic to a subgroup of $\mc{U}_\mf{F}^\Omega$ through its unitary representation on $\mc{H}_\Omega$. Then, a \emph{continuous physical symmetry} is given by a (strongly) continuous one-parameter group of automorphisms $\alpha\colon s\mapsto \alpha_s \in \mt{Aut}(\mf{A}_\Omega)$ that are induced by some unitaries in $\mc{U}_\mf{F}^\Omega$. By Stone's theorem, the group of unitaries $s \mapsto U_s$ representing the continuous symmetry $\alpha$ on $\mc{H}_\Omega$ is generated as $U_s = e^{isL_\alpha}$ by a (possibly unbounded but densely def\/ined) self-adjoint operator $L_\alpha$ af\/f\/iliated with $B(\mc{H}_\Omega)$ that satisf\/ies $L_\alpha |\1_\mf{F}\rangle_\Omega = 0$ \cite{KadisonRingrose,KadisonRingrose2}.

\subsection{Equilibrium condition and thermal dynamics}\label{subsec:modular}
The reference state $\tilde\Omega\in\mc{S}(\mf{F})$ is faithful on $\pi_\Omega(\mf{F})$ if the null ideal $\mc{N}_\Omega = \{a\in \mf{F} \colon \Omega(a^*a) = 0\}$ satisf\/ies $\mc{N}_\Omega = \ker(\pi_\Omega) := \{a\in\mf{F} \colon \pi_\Omega(a) = 0\}$. This is equivalent with $\mc{N}_\Omega$ being two-sided and therefore self-adjoint (i.e., if $\Omega(a^*a) = 0$ for $a\in\mf{F}$, then also $\Omega(aa^*) = 0$). In terms of measurement probabilities the self-adjointness of $\mc{N}_\Omega$ corresponds to the following statistical property of the reference state: If any sequence of spectral projections $P_{O_1}^{(i_1)} P_{O_2}^{(i_2)}\cdots P_{O_n}^{(i_n)}$ has a vanishing probability via equation~\eqref{eq:probs}, then also the probability for the reversed sequence $P_{O_n}^{(i_n)} P_{O_{n-1}}^{(i_{n-1})}\cdots P_{O_1}^{(i_1)}$ vanishes. Accordingly, the requirement $\mc{N}_\Omega = \ker(\pi_\Omega)$ is implied by the detailed balance condition, which states that the probabilities for any process and its reverse are the same in equilibrium, and actually implies that~$\Omega$ may represent an equilibrium state.

In particular, when the extended reference state $\tilde\Omega$ is faithful on the physical observable algebra $\mf{A}_\Omega := \pi_\Omega(\mf{F})'' \subset B(\mc{H}_\Omega)$, it gives rise to the following two canonical operators on the GNS Hilbert space $\mc{H}_\Omega$ by Tomita--Takesaki modular theory \cite{Summers06, Takesaki2}:
\begin{enumerate}\itemsep=0pt
	\item The \emph{modular operator} $\Delta_\Omega$ is a (possibly unbounded) positive operator af\/f\/iliated with $B(\mc{H}_\Omega)$, which satisf\/ies $\Delta_\Omega |\1_\mf{F}\rangle_\Omega = |\1_\mf{F}\rangle_\Omega$ and $\Delta_\Omega^{it}\mf{A}_\Omega \Delta_\Omega^{-it} = \mf{A}_\Omega$ for all $t\in\mb{R}$. The modular operator induces a strongly continuous one-parameter group of automorphisms of $\mf{A}_\Omega$, the \emph{modular flow}, $\sigma^\Omega\colon t \mapsto\sigma^\Omega_t \in \mt{Aut}(\mf{A}_\Omega)$ through the unitary action
	\begin{gather*}
		\Delta_\Omega^{it} a \Delta_\Omega^{-it} \equiv \sigma^\Omega_t(a) \qquad \forall\, a\in\mf{A}_\Omega,\ t\in\mb{R} .
	\end{gather*}
	The extended reference state $\tilde\Omega$ satisf\/ies the Kubo--Martin--Schwinger (KMS) equilibrium condition with respect to the modular f\/low $\sigma^\Omega$, and the modular f\/low is the unique one-parameter group of automorphisms of $\mf{A}_\Omega$ with this property, up to rescaling $t \mapsto \lambda t$, $\lambda\in \mb{R}_+$ of the f\/low parameter \cite{BratteliRobinson87,BratteliRobinson87+}. (This rescaling freedom means that the temperature of the equilibrium state is indetermined.)
	\item The \emph{modular involution} $J_\Omega$ is an anti-linear involutive operator (i.e., $J_\Omega^* = J_\Omega^{-1} = J_\Omega$), which satisf\/ies $J_\Omega |\1_\mf{F}\rangle_\Omega = |\1_\mf{F}\rangle_\Omega$ and $J_\Omega \mf{A}_\Omega J_\Omega = \mf{A}_\Omega'$. Accordingly, $\mf{A}_\Omega$ and $\mf{A}_\Omega'$ are (anti) isomorphic von Neumann algebras. Moreover, we have $J_\Omega \Delta_\Omega J_\Omega = \Delta_\Omega^{-1}$.
\end{enumerate}
These two operators arise from the polar decomposition of the operator $S_\Omega \equiv J_\Omega \Delta_\Omega^{\frac{1}{2}} \colon \mc{H}_\Omega \rightarrow \mc{H}_\Omega$ def\/ined through its action $S_\Omega |a\rangle_\Omega = |a^*\rangle_\Omega$ for all $a\in\mf{F}$, and are therefore completely canonical. We may also consider the \emph{modular generator} $D_\Omega := -\ln \Delta_\Omega$, which is a self-adjoint operator af\/f\/iliated with $B(\mc{H}_\Omega)$. The modular generator annihilates the cyclic vector, $D_\Omega |\1_\mf{F}\rangle_\Omega = 0$, and therefore generates a continuous physical symmetry. In particular, $D_\Omega$ generates the one-parameter group represented by the unitaries $\Delta_\Omega^{it} \equiv e^{-itD_\Omega}$. The spectrum of $D_\Omega$ is always symmetric with respect to zero, and therefore it cannot be directly interpreted as the Hamiltonian of the system. For the GNS representation induced by a thermal state of a f\/inite-dimensional quantum system, $D_\Omega$ in fact corresponds to the Liouville operator $L_H(A) \equiv [H,A]$, where $H$ is the Hamiltonian of the system. On the other hand, $D_\Omega$ is not always af\/f\/iliated with the observable algebra for inf\/inite-dimensional systems, because the unitaries $\Delta_\Omega^{it} \in B(\mc{H}_\Omega)$ do not belong to $\mf{A}_\Omega$ for all $t\in\mb{R}$ (i.e., they induce outer automorphisms of $\mf{A}_\Omega$), and therefore cannot be approximated by physical measurements.\footnote{Note that this is, however, in a qualitative agreement with general relativity, where there does not exist a~general globally def\/ined observable for the total energy of a system \cite{Wald84}.}

Connes and Rovelli \cite{ConnesRovelli94} have suggested to consider the one-parameter group of automorphisms given by the modular f\/low $\sigma^\Omega\colon \mb{R} \rightarrow \mt{Aut}(\mf{A}_\Omega)$ as the physical time-evolution for a background-independent quantum system~-- the so-called \emph{thermal time hypothesis}. Indeed, $\sigma^\Omega$ gives the unique dynamics on $\mf{A}_\Omega$ with respect to which the extended reference state $\tilde\Omega$ is in equilibrium. We will apply this idea in the spacetime-free framework in order to recover `thermal' unitary dynamics for a quantum system in the case that the extended reference state $\tilde\Omega$ may represent equilibrium, i.e., when it is faithful on the physical observable algebra $\mf{A}_\Omega$.

There are a few apparent challenges to the thermal time hypothesis: (1) To begin with, a~pure state does not induce a non-trivial modular structure, and therefore the vacuum state of QFT cannot give rise to global dynamics. This problem can be overcome by noting that the restriction of the vacuum onto a subregion of spacetime gives rise to a thermal state that does induce non-trivial modular dynamics. For example, the restriction of the Minkowski vacuum of a~free neutral scalar f\/ield onto a half-space is well-known (by the Bisognano--Wichmann theorem) to give rise to a modular f\/low that is given by the one-parameter group of Lorentz boosts that preserve the corresponding Rindler wedge~\cite{Borchers00}. In this case, the integral curves of the f\/low correspond to the worldlines of accelerated observers, for whom the boundary of the Rindler wedge is an apparent causal horizon. Correspondingly, the modular generator is the generator of the proper time evolution of these observers, so the modular f\/low is indeed seen to give the time-evolution of a particular class of observers. In the universe we inhabit, no localized observer can access all the degrees of freedom in the universe, but only her causal past (the `observable universe'), so the restriction of the global pure state even for an inertial observer can be physically justif\/ied in this way. In addition, our universe is f\/illed with thermal cosmic background radiation. Remarkably, Rovelli~\cite{Rovelli93} has shown that the thermal dynamics induced by the statistical state describing cosmic background radiation in the Robertson--Walker spacetime agrees with the usual cosmological dynamics with respect to the Robertson--Walker time. (2)~Secondly, for QFT on curved spacetime there do not exist any equilibrium states, unless the background spacetime is static~\cite{Solveen12, Wald94}, implying that for the majority of spacetimes we cannot obtain the spacetime structure from a thermal state. Actually, this seemingly problematic point is consistent with our point of view, according to which the spacetime structure is determined by the reference state: Clearly, for the ef\/fective spacetime geometry to be static, the reference state must be suitably invariant under the dynamics.\footnote{The exact form of the invariance depends on the way that the ef\/fective spacetime geometry depends on the reference state.} However, the spacetime-free framework also applies to the case where the reference state cannot represent equilibrium, although in this case we cannot recover thermal dynamics, and we must use other properties of the system to extract the ef\/fective spacetime geometry. (3)~Thirdly, the modular f\/low for QFT states does not in fact correspond generically to the time-evolution of the system~\cite{Borchers00}. In a few cases, such as the vacuum state restricted to a Rindler wedge in Minkowski spacetime, the modular f\/low is seen to be related to time-evolution, but in most known cases the action of the modular f\/low is non-local with respect to the background geometry. We might interpret this as signaling the incompatibility of such a state to act as a reference state for the particular background spacetime geometry, as in the spacetime-free formulation the state should probably give rise to a dif\/ferent ef\/fective background geometry (if any), with respect to which the dynamics are local. In fact, in Section~\ref{sec:spacetime} we will explore the idea that a notion of locality for the ef\/fective spacetime structure can be \emph{defined} by the requirement that the thermal dynamics induced by the reference state are local.

Finally, let us also mention that the modular involution $J_\Omega$ for the vacuum state restricted to Rindler wedges in Minkowski spacetime is known to have a physical interpretation as a~combination of the CPT operator and a~spatial rotation for relativistic QFT models that satisfy certain general algebraic requirements \cite{Borchers00}. Accordingly, the existence of the modular involution is strongly related to Lorentz invariance through the CPT theorem and the symmetry between matter and antimatter (i.e., retarded and advanced solutions to the equations of motion), since the CPT transformation in QFT maps particles to antiparticles, and vice versa~\cite{Haag96}. Indeed, taking advantage of the relation between the modular involutions and CPT transformations, it has been shown in \cite{Buchholz98,SummersWhite03,WhiteThesis} that the modular involutions induced by the QFT vacuum restricted to Rindler wedges can be used to generate a representation of the proper Poincar\'e group, thus recovering spacetime structure from the purely algebraic data of a vacuum state and a suitable family of subalgebras of observables. However, it is unclear whether this method for deriving spacetime from algebraic data can be extended to less symmetric situations.

\subsection{Perturbations of the reference state}\label{subsec:perturbations}
The extended reference state $\tilde\Omega$ may be perturbed by a f\/inite collection of operators $b_k\in B(\mc{H}_\Omega)$ by def\/ining the perturbed reference state as
\begin{gather*}
	\tilde\Omega'(a) := N^{-1}\sum_k \tilde\Omega(b_k^*ab_k)
\end{gather*}
for all $a\in\mf{A}_\Omega$, where $N := \sum_k \tilde\Omega(b_k^*b_k) \in \mb{R}_+$ is a normalization constant (assumed to be nonzero). Such a perturbed state lies in the folium of the reference state as it is represented by the density operator\footnote{The folium of a state $\omega$ consists of those states, which can be represented by density operators in $B(\mc{H}_\omega)$~\cite{Haag96}.}
\begin{gather*}
	\rho_{\tilde\Omega'} := N^{-1} \sum_k b_k |\1_\mf{F}\rangle_\Omega \langle \1_\mf{F}| b_k^* \in B(\mc{H}_\Omega) .
\end{gather*}
The perturbed extended reference state $\tilde\Omega'$ gives rise to a perturbed state $\Omega'$ on the free observable algebra $\mf{F}$ by its restriction onto $\pi_\Omega(\mf{F}) \subset B(\mc{H}_\Omega)$.

We may then consider the GNS representation $\pi_{\Omega'}\colon \mf{F} \rightarrow B(\mc{H}_{\Omega'})$ of $\mf{F}$ induced by such a~perturbed reference state that gives rise to the perturbed physical observable algebra \mbox{$\mf{A}_{\Omega'}\! :=\! \pi_{\Omega'}(\mf{F})''$}. In this way, perturbations of the state of the system may af\/fect the algebraic as well as the statistical relations of the physical observables, which we suspect may be associated with the change of the ef\/fective spacetime structure and thus gravitational ef\/fects. Importantly, we always have $\ker\pi_\Omega \subset \ker\pi_{\Omega'}$ for perturbations of $\tilde\Omega$ by elements in $B(\mc{H}_\Omega)$, which implies that such perturbations cannot destroy the joint measurability of observables in $\mf{F}$: If $[a,a'] \in \ker\pi_\Omega$ for some $a,a'\in\mf{F}$, then also $[a,a'] \in \ker\pi_{\Omega'}$. This is very important from the physical point of view, because otherwise perturbations to the system could completely change the physical properties (e.g., the causality structure) of the system. Notice that $\ker\pi_\Omega \subset \ker\pi_{\Omega'}$ can be a~proper inclusion only if $\pi_\Omega$ is reducible. Also, the perturbed state $\tilde\Omega'$ cannot be faithful on the unperturbed physical observable algebra~$\mf{A}_\Omega$. (However, $\tilde\Omega'$ may still be faithful on $\mf{A}_{\Omega'}$.) We take the physical states of the original system to inhabit the folium of the reference state. Accordingly, we will call $B(\mc{H}_\Omega)$ the \emph{perturbation algebra}, as it induces perturbations of the reference state. In ordinary QFT, the quantum f\/ield operators belong to the perturbation algebra.\footnote{Actually, the smeared f\/ield operators in QFT should correspond to unbounded operators af\/f\/iliated with~$B(\mc{H}_\Omega)$, but we may consider the exponentiated (Weyl) f\/ield operators, which are bounded. The original unbounded f\/ield operators are recovered as generators of strongly continuous one-parameter groups of unitaries in~$B(\mc{H}_\Omega)$. See \cite{Haag96,HalvorsonMueger07} for the algebraic construction of quantum f\/ields from the representations of the observable algebra.} As for the f\/ield operator algebra in algebraic QFT, the perturbation algebra provides an extension of the physical observable algebra.

The action of any physical symmetry implemented by a unitary $U \in \mc{U}_\mf{F}^\Omega$ on the density operators as $\rho_{\tilde\Omega'}\mapsto U \rho_{\tilde\Omega'} U^* \in B(\mc{H}_\Omega)$ maps the folium of the reference state to itself. Therefore, the perturbations carry a representation theory of the group of physical symmetries (although it may very complicated in general).

In Appendix~\ref{app:positiveenergy} we propose a def\/inition for the mass of a static perturbation (based on the Connes cocycle derivative), and argue that perturbations whose mass is positive according to this def\/inition always render some new observables jointly measurable (i.e., mutually commutative), which we suggest is reminicient of lightcone focusing by gravity. We have delegated this rather technical and provisional discussion to an appendix in order to make the presentation more streamlined for the benef\/it of the reader.

This f\/inishes our sketch of a spacetime-free framework for quantum physics. In the next section we will consider some ways to recover information about the possible ef\/fective spacetime structure induced by the reference state and its perturbations. However, before that we will brief\/ly discuss the connection of the spacetime-free framework to the usual (algebraic) formulation of quantum f\/ield theory. Also, for the simplest concrete examples of the above construction, see Appendix \ref{app:examples}.

\subsection{Relation of the spacetime-free framework to quantum f\/ield theory}
The above formulation of the spacetime-free framework in terms of the free observable algebra is rather reminicent of the construction of the Borchers--Uhlmann algebra in algebraic quantum f\/ield theory \cite{Borchers75}. In short, the Borchers--Uhlmann algebra is obtained as the free tensor algebra over the vector space of Schwarz functions on Minkowski spacetime. Then, a state specif\/ied by the Wightman functionals imposes non-trivial algebraic relations between the elements of the Borchers--Uhlmann algebra, which encode the dynamics of the model. There are at least two important dif\/ferences between the two constructions:
\begin{itemize}\itemsep=0pt
	\item Unlike the free observable algebra in the spacetime-free framework, the Borchers--Uhlmann algebra is constructed on top of a f\/ixed background spacetime geometry, which is obviously antithetical to the goals of our approach.
	\item In the case of the Borchers--Uhlmann algebra, the elements of the free tensor product are functions in spacetime, whereas in our case we constructed the free product of abelian von Neumann algebras generated by self-adjoint operators. This ref\/lects a dif\/ference in the physical interpretation of the elements of the construction. Namely, the elements of the Borchers--Uhlmann algebra are not necessarily physical observables, but correspond to localized f\/ields in the absence of dynamical relations, i.e., they are `kinematical' operators. In our case, the requirement that the initial algebras are faithfully represented through the GNS representation implies that they correspond to physical observables.
\end{itemize}
The common key idea between the two formalisms is, however, that the quantum state imposes dynamical relations on the algebra of observables.
Indeed, in some cases the formalisms seem to be closely related. In particular, we could restrict to consider some subset of operators of the Borchers--Uhlmann algebra that is generated by functions localized to Cauchy surfaces, so that they are in fact physical observables. Then, the Wightman functionals would provide the reference state encoding the statistics of observations for these observables. However, the exact details of the relationship remain to be worked out.

We would also like to point out one possible confusion that may arise from quantum f\/ield theo\-ry concerning the choice of the reference state. In ordinary QFT most states on the observable algebra are considered unphysical~-- for example, those not satisfying the Hadamard condition~\cite{BarFredenhagen09}. However, such criteria for physical states usually rely on the background spacetime structure, and in ef\/fect guarantee that the state is compatible in some way with the f\/ixed background geometry. In the absence of a background geometry, however, these criteria are inapplicable, so how are we able to distinguish the physical states from the unphysical ones? We would like to point out that a state that appears wildly unphysical with respect to some f\/ixed background geometry, may in fact be well-behaved with respect to the ef\/fective background geometry, to which it gives rise. On the other hand, when the background spacetime is not restricting the symmetries of the system, the equivalence classes of states describing the same physics should be much larger than with a f\/ixed background spacetime. In particular, as we observed above, any covariant transformation $(\mf{F},\Omega) \mapsto (\alpha(\mf{F}), \alpha^*\Omega)$ given by a $*$-automorphism $\alpha\in\mt{Aut}(\mf{F})$ leads to the same physical description of the quantum system in the spacetime-free formulation, whereas if the observables were labeled by some form of background spacetime information (e.g., spacetime regions) from the beginning, such transformations would not in general correspond to geometric transformations of the background spacetime structure. Therefore, the physics described by the spacetime-free framework may be more unique than that of ordinary QFT.

\section{Recovering ef\/fective spacetime structure}\label{sec:spacetime}
In this section we will develop some ideas and methods for the extraction of spacetime structure from the spacetime-free framework, which is clearly necessary in order to connect the theory with known physics and experiments. In fact, the reconstruction of spacetime has been considered before in the context of algebraic quantum f\/ield theory in the literature. The earliest works we have found addressing the issue are~\cite{Bannier94,Keyl98}, where the inverse problem of recovering spacetime topology and causal structure from the net of local observable algebras is considered. On the other hand, \cite{Buchholz98, SummersWhite03, WhiteThesis} show that it is possible to recover the symmetry group of spacetime from the modular structure of subalgebras in some highly symmetric cases. These results already show that quite a lot of information about spacetime structure is encoded into quantum f\/ield theory. But they also carry the contradiction within them that the original formulation of quantum f\/ield theory relies on a background spacetime, which is what we try to remedy in this work.

\subsection{Locality from the dynamical properties of subalgebras}
By locality we refer to the existence and the identif\/ication of local subsystems. We may distinguish the following two notions of locality that are \emph{a priori} independent\footnote{Similar distinction between notions of locality has been made before, e.g., in~\cite{FewsterVerch11}.}:
\begin{itemize}\itemsep=0pt
	\item \emph{Locality with respect to a background geometry.} The topology of the background manifold gives rise to a geometrical notion of local spacetime regions and the corresponding localized subsystems in the usual formulations of QFT.
	\item \emph{Locality with respect to the dynamics.} The dynamics may give rise to an operational notion of locality, which can be determined by studying the evolution of matter systems, e.g., the propagation of excitations over the vacuum.
\end{itemize}
The choice of dynamics in f\/ield theory is usually guided by the principle of locality, which can be understood as the requirement that local subsystems interact with the rest of the system only at the boundary of the corresponding local spacetime region. The requirement of locality for interactions with respect to the background geometry ensures that the dynamical notion of locality agrees with the geometrical notion of locality. In the absence of a background geometry, one must solely rely on the dynamical notion of locality, and thus base the def\/inition of locality on the dynamical properties of subsystems. In particular, we suggest to study the propagation of causal inf\/luences as encoded into the commutation relations between observables.\footnote{On the other hand, let us emphasize that there may also exist more algebraic procedures to identify local subalgebras of observables, such as the method of modular localization~\cite{BrunettiGuidoLongo02} for free quantum f\/ield theory on Minkowski spacetime. However, modular localization relies heavily on the symmetry properties of spacetime and the Fock space structure, and therefore is not directly applicable to generic spacetimes or interacting theories.}

In algebraic QFT, observable algebras $\mf{A}(\mc{O})$ are associated to spacetime regions $\mc{O}$. Typically, the observable algebras associated with local spacetime regions are von Neumann subfactors (i.e., $\mf{A}(\mc{O})'' = \mf{A}(\mc{O})$ and $\mf{A}(\mc{O})\cap \mf{A}(\mc{O})' \cong \mb{C}$) of the total observable algebra. Moreover, many of the models satisfy the \emph{Haag duality}: $\mf{A}(\mc{O}_c) = \mf{A}(\mc{O})'$, where $\mc{O}_c$ denotes the causal complement to the region~$\mc{O}$~\cite{Haag96}. These properties can be physically motivated by the joint measurability of commuting observables. Specif\/ically, the commutant $\mf{A}(\mc{O})'$ contains all the observables that are jointly measurable with all the observables in $\mf{A}(\mc{O})$ and, therefore, describes the rest of the degrees of freedom in addition to those in $\mf{A}(\mc{O})$ that are necessary to completely specify the quantum state of the system. As an example, in QFT for the free scalar f\/ield, the f\/ield values on any Cauchy surface determine a pure state of the system, as the corresponding f\/ield operators form a maximal abelian subalgebra of the physical observable algebra, i.e., a complete set of jointly measurable observables. We may then think of $\mf{A}(\mc{O})$ and $\mf{A}(\mc{O})'$ as splitting some Cauchy surface in two parts, as any maximal abelian subalgebra is correspondingly split between the two algebras, each of them describing the degrees of freedom associated to one of the two complementary parts of the Cauchy surface. As the joint measurability of observables and the maximal abelian subalgebras retain their physical interpretation even in the absence of a~background spacetime, we will require any subalgebra of observables $\mf{B} \subset \mf{A}_\Omega$ corresponding to a~subsystem to be a von Neumann subfactor, and we take the relative commutant $\mf{B}' = \{b'\in \mf{A}_\Omega \colon [b,b'] = 0\, \forall\, b\in \mf{B}\}$ to represent the causal complement, or the environment, to the subsystem represented by~$\mf{B}$.

The propagation of causal inf\/luences manifests in how the thermal dynamics induced by the modular f\/low $\sigma^\Omega\colon \mb{R} \rightarrow \mt{Aut}(\mf{A}_\Omega)$ af\/fects the commutators between dif\/ferent subalgebras of observables. In particular, the evolution $\mf{B} \mapsto \sigma^\Omega_t(\mf{B})$ tends to `scramble' the local subalgebras of observables $\mf{B}\subset \mf{A}_\Omega$. Let $\mf{B}_1,\mf{B}_2 \subset \mf{A}_\Omega$ be two subfactors such that $\mf{B}_1 \subset \mf{B}_2'$. We may then consider the commutators $[\sigma^\Omega_t(b_1), b_2]$ for dif\/ferent $t \in\mb{R}$ and $b_1\in\mf{B}_1$, $b_2\in\mf{B}_2$, and their (normalized weak operator) norms
\begin{gather*}
	\frac{\|[\sigma^\Omega_t(b_1), b_2]\|}{2\|b_1\| \|b_2\|} \in [0,1] .
\end{gather*}
Clearly, at $t = 0$ the norm vanishes, and its growth away from $t=0$ for some elements $b_1\in\mf{B}_1$ and $b_2\in\mf{B}_2$ indicates the necessity of some causal relations forming between the degrees of freedom associated to the two subfactors. In local QFT, for two f\/ield subalgebras that are local and f\/initely spacelike separated, the commutator for any pair of elements remains zero for a f\/inite interval around $t=0$, if the Einstein causality property is satisf\/ied (i.e., propagation happens inside lightcones).\footnote{A model in algebraic QFT is said to satisfy Einstein causality if $\mf{B}_1 \subset \mf{B}_2'$ whenever $\mf{B}_1$ and $\mf{B}_2$ are subalgebras associated with spacelike separated spacetime regions \cite{Haag96}.} Even if Einstein causality is violated by the model (e.g., for non-relativistic systems such as spin chains), the derivative
\begin{gather*}
	\frac{\dd}{\dd t} [\sigma^\Omega_t(b_1), b_2] = i[[D_\Omega, b_1], b_2]
\end{gather*}
still vanishes, if the thermal dynamics does not directly couple $\mf{B}_1$ and $\mf{B}_2$. Therefore, we see that the behavior of the quantity
\begin{gather*}%\label{eq:correlation}
	C^\Omega_t(\mf{B}_1, \mf{B}_2) := \sup \left\{ \frac{\|[\sigma^\Omega_t(b_1), b_2]\|}{2\|b_1\| \|b_2\|} \in [0,1] \colon b_1 \in \mf{B}_1,\ b_2 \in \mf{B}_2 \right\} \in [0,1]
\end{gather*}
as a function of $t$ encodes information on the dynamical coupling of subfactors $\mf{B}_1$ and $\mf{B}_2$. In particular, its growth can in some cases be used to estimate the strength of direct dynamical coupling between~$\mf{B}_1$ and~$\mf{B}_2$. Our primary suggestion for a strategy to recover the operational topology of the experimental system is based on studying the magnitude and growth of such commutators to estimate the strength of causal dependence between dif\/ferent subsystems.

It is possible that for some relativistic inf\/inite-dimensional systems $C^\Omega_t(\mf{B}_1, \mf{B}_2)$ only takes values~0 and~1. In this case, we may extract information on the spatial distance between subsystems from the value of the parameter $t$, for which the value of $C^\Omega_t(\mf{B}_1, \mf{B}_2)$ changes from~0 to~1. In the following, however, we will consider the case that $C^\Omega_t(\mf{B}_1, \mf{B}_2)$ behaves continuously in $t$ (e.g., for f\/inite-dimensional systems). In this case, we would like to identify the local subsystems as those represented by subfactors $\mf{B} \subset \mf{A}_\Omega$, which are most weakly coupled to their commutant~$\mf{B}'$, representing the causal complement to the subsystem. This def\/inition is motivated by the principle of locality as interpreted above: Local subsystems interact with their environment only through their boundary. In this sense, local subsystems are the most robust against inf\/luences from their environment. The strength of the dynamical coupling between a subsystem represented by~$\mf{B}$ and its causal complement~$\mf{B}'$ can be estimated by the growth of~$C^\Omega_t(\mf{B},\mf{B}')$ at $t=0$. In particular, we expect $\frac{\dd}{\dd t}C_t^\Omega(\mf{B},\mf{B}')$ at $t=0$ to be minimized by the subset of \emph{local spherical} subfactors in any set of subfactors $\{\alpha(\mf{B}_0) \subset \mf{A}_\Omega \colon \mf{B}_0 \subset \mf{A}_\Omega,\, \alpha \in \mt{Aut}(\mf{A}_\Omega)\}$ connected by automorphisms of~$\mf{A}_\Omega$. By the term `spherical' we take into account that the magnitude of $\frac{\dd}{\dd t}C_t^\Omega(\mf{B},\mf{B}')$ may also depend on the size of the boundary of the spatial region associated with~$\mf{B}$. The totality of local subfactors can then be obtained as the net of subfactors generated by the local spherical ones. The local subfactors are determined only up to the symmetry group of the reference state, since $C_t^\Omega(\mf{B},\mf{B}')$ is invariant under symmetry transformations. In other words, a local subfactor is mapped to a~local subfactor by the symmetry transformations.

Ultimately, to justify the above def\/inition of local subsystems, we should be able to show that the growth of $C^\Omega_t(\mf{B},\mf{B}')$ (or a quantity similar to it) is indeed minimized by a class of local subsystems in the usual formulation of local QFT (or some f\/inite-dimensional regularization thereof). Unfortunately, $C^\Omega_t(\mf{B},\mf{B}')$ is rather dif\/f\/icult to compute explicitly due to the supremum over algebra elements, and thus this def\/inition remains largely a hypothesis or a suggestion for now. However, in Appendix \ref{app:spinlattices} we study the behavior of a similar quantity measuring the magnitude of commutators for f\/inite spin lattices, which is straighforward to compute. We have verif\/ied that the growth of this quantity is indeed minimized for certain local subalgebras, which of\/fers at least some plausibility for the above def\/inition. Nevertheless, the full verif\/ication of the hypothesis is left for future work.

\subsection{Metric information from the readings of quantum clocks}
In the previous subsection, we proposed how to identify the local perturbation subalgebras, and thus def\/ine a notion of locality for the system. In this subsection, we will brief\/ly consider some preliminary ideas on how to recover metric data. As the starting point of our considerations, we take the view that the operational geometry is determined by the readings of clocks.\footnote{The recent work~\cite{CaoCarrollMichalakis16} by Cao, Carroll and Michalakis nicely complements our approach by showing that one can reconstruct spatial metric information from the mutual information shared by dif\/ferent subsystems.}

Quantum clock systems have been considered before mainly in the context of non-relativistic quantum mechanics (see, e.g., \cite{Gambini07, Peres80,Salecker58, Tralle05}). It was found that quantum ef\/fects impose inherent restrictions on the accuracy of measurements of duration and distance. Our treatment of quantum clocks dif\/fers quite signif\/icantly from the earlier works, since we work in the spacetime-free framework for quantum physics as formulated in Section \ref{sec:framework}. Nevertheless, similar restrictions are expected to be valid due to uncertainty relations, which still arise from the non-commutativity of observables.

Let us also mention, although we will not explore this option any further here, that it may be possible to apply non-commutative geometric methods \`a la Connes to recover metric information \cite{Connes94,Varilly06}. Indeed, the modular structure of von Neumann algebras is remarkably similar to the def\/inition of a spectral triple, which is a natural non-commutative generalization of a metric space \cite{BertozziniContiLewkeeratiyutkul08,BertozziniContiLewkeeratiyutkul10}. However, the modular generator, unlike the usual Dirac operator, operates in the `quantum phase space' and not on spacetime, \emph{per se}. Therefore, we expect the spectral geometry induced by the modular structure to describe the non-commutative phase space geometry rather than directly spacetime. The relation between the two can be quite complicated in f\/ield theory. Accordingly, we consider it a more feasible f\/irst take on the problem of obtaining spacetime metric data to study the evolution of quantum clock systems.

Let us again assume that the reference state $\tilde\Omega$ is faithful on the physical observable algebra~$\mf{A}_\Omega$, and therefore gives rise to the thermal dynamics. Since there is no physical evolution in equilibrium, we must consider perturbations to the reference state. A perturbation $\tilde\Omega'$ of the reference state $\tilde\Omega$ is localized to a local subsystem represented by a local observable subalgebra $\mf{B}\subset \mf{A}_\Omega$, if it satisf\/ies $\tilde\Omega'(b') = \tilde\Omega(b')$ for all $b'\in \mf{B}'$. A simple clock system can be taken to consist of a (possibly approximately) local perturbation $\tilde\Omega'$ of the reference state and an associated observable (i.e., a self-adjoint operator) $\gamma\in\mf{A}_{\Omega'}$ representing the reading of the clock. The evolution of a perturbation $\tilde\Omega'$ relative to the reference state $\tilde\Omega$ is given by the modular f\/low $\sigma^\Omega\colon \mb{R} \rightarrow \mt{Aut}(\mf{A}_\Omega)$ as explained in Section~\ref{subsec:modular}. We may then require that the expectation value
\begin{gather*}
	\langle \gamma \rangle_{\Omega',t} := \tilde\Omega'\big(\sigma^\Omega_t(\gamma)\big) \in \mb{R}
\end{gather*}
of the clock reading increases monotonically during the evolution of the system, which suggests the requirement $ i\tilde\Omega'\big([D_\Omega, \gamma] \big) > 0$ for the pair $(\tilde\Omega', \gamma)$ specifying the clock. If such an observable~$\gamma$ is found, we may always further scale $\gamma \mapsto r\gamma$ by a positive real number $r\in\mb{R}_+$, so that $ i\tilde\Omega'\big([D_\Omega, \gamma] \big) = 1$. Adopting the relativistic terminology, we may say that such an observable $\gamma$ measures the \emph{proper time} of the perturbation $\tilde\Omega'$. For the clock to be of practical use, we should also require the distribution of the value of $\gamma$ to be peaked around its expectation value.

Now, let $\mf{B}_1,\mf{B}_2\subset B(\mc{H}_\Omega)$ be two subalgebras of perturbations localized in dif\/ferent local subsystems. To measure the proper time that it takes for a clock to travel between the two subsystems, we must f\/ind a clock system that propagates from one to the other. Let $\tilde\Omega'$ be a perturbation induced from the reference state $\tilde\Omega$ by an operator $b\in B(\mc{H}_\Omega)$ such that $\Delta_\Omega^{-it_1} b \Delta_\Omega^{it_1} \in\mf{B}_1$ and $\Delta_\Omega^{-it_2} b \Delta_\Omega^{it_2} \in\mf{B}_2$ for some $t_1,t_2\in\mb{R}$, $t_1<t_2$. Then, we could def\/ine for the clock pair $(\tilde\Omega', \gamma)$ the expectation value of the measured time between the two local spacetime regions as the dif\/ference $\tilde\Omega'\big(\sigma^\Omega_{t_2}(\gamma) - \sigma^\Omega_{t_1}(\gamma) \big)$ in the clock readings at the two times.

We suggest that it is possible to recover the ef\/fective metric relationships between local spacetime regions def\/ined by the local perturbation subalgebras by studying the evolution of such quantum clock systems. Of course, in general, one must use several clock systems simultaneously, in which case one needs to f\/ind several local perturbation operators $b_i\in B(\mc{H}_\Omega)$, $i=1,\ldots,n$ for some $n\in\mb{N}$, which evolve approximately independently from each other, propagating between local subalgebras of $B(\mc{H}_\Omega)$, and a corresponding set of (approximately) mutually commuting observables $\gamma_i\in\mf{A}_{\Omega'}$, $i=1,\ldots,n$, where $\tilde\Omega'$ is the reference state perturbed by each of \mbox{$b_i\in B(\mc{H}_\Omega)$}. Finding a family of such perturbations, which determines the spacetime geometry to a~satisfying accuracy, is undoubtedly a highly non-trivial task. Whether such families of quantum clock systems can actually be found for some systems remains an open theoretical question at the moment, although it appears to us that we use exactly such systems to determine spacetime structure in practice. We will leave the further development of these ideas to future work.

\subsection{Perturbations of the ef\/fective spacetime structure and gravity}
Finally, we wish to mention a couple of ideas concerning the relationship between the perturbations of the reference state and gravitational phenomena. We saw in Section \ref{sec:framework} how perturbations of the reference state may change the causal relations of physical observables by altering the GNS representation of the free observable algebra. In Section \ref{sec:framework} (and Appendix \ref{app:positiveenergy}) we argued that perturbations with positive mass always causally decouple some observables, which seems at least suggestive of gravitational phenomena. It is evident that perturbations of the reference state also alter the ef\/fective geometric properties of the quantum system as def\/ined in this section, but we have not studied the exact nature of such perturbations so far.

Let us also mention the recent work \cite{Jacobson15} by Jacobson, in which he derives the semiclassical Einstein equation from the hypothesis that the QFT vacuum state restricted locally to small causal diamonds maximizes the (von Neumann) entanglement entropy. The derivation points to another way of relating perturbations of the reference state (in this case the vacuum) to gravity in terms of quantum statistics instead of representation theory. However, in Jacobson's derivation the division of entropy into UV (high energy) and IR (low energy) parts plays an important role: The entropy associated to the UV degrees of freedom leads via the area law to the geometric part of the Einstein equation, while the IR entropy is related to the matter energy-momentum tensor. It is argued that the simultaneous perturbation of the two vanishes for local subsystems due to the maximization of the total entropy by the vacuum state, and out comes the Einstein equation. In our formalism it is not clear how to divide the degrees of freedom into the UV and IR parts. In the presence of an equilibrium state we do have a notion of energy provided by the eigenvalues of the modular operator, but the observable algebra does not generally factorize into high and low energy parts. We could require of the reference state for such factorization to hold, so that low and high energy degrees of freedom are (approximately) decoupled. However, we are also able to do without such an assumption if the total entanglement entropy for the restriction of the vacuum onto local subsystems satisf\/ies the area law, i.e., its leading contribution is proportional to the area of the boundary of the local subsystem\footnote{The entanglement entropy is usually divergent in QFT. To make it f\/inite, some kind of a regularization must be introduced.}: The `f\/irst law of entanglement entropy' \cite{Blanco13} implies for f\/irst order variations of a thermal equilibrium state that $\delta S = \delta \langle H_\Omega \rangle$, where $S$ is the entanglement entropy and $\langle H_\Omega \rangle$ is the expectation value of the (modular) Hamiltonian induced by the reference state $\Omega$.\footnote{Here, the modular Hamiltonian is obtained as the generator of the modular f\/low when the modular f\/low is given by inner automorphisms, i.e., can be induced by the adjoint action of unitaries in the physical observable algebra. We are not aware of the extension of the entanglement f\/irst law to the general case, where the generator of the time-evolution does not belong to the observable algebra.} From the entanglement f\/irst law we may derive the semiclassical (linearized) Einstein equation for ball shaped spatial regions by relating the entropy to the spacetime geometry via the area law and the modular Hamiltonian to the energy-momentum tensor exactly as in the derivation \cite{Jacobson15} of Jacobson. This idea has also been considered before in the context of AdS/CFT correspondence (see, e.g., \cite{Lashkari13}). To apply it in the spacetime-free framework, we should require that the reference state is in thermal equilibrium when restricted to local subsystems. Then, if we can relate spacetime geometry to the entanglement entropy via the area law, this derivation shows that Einstein gravity could naturally arise from the quantum statistics, at least in its linearized form.

Unfortunately, this is as far as our understanding of the exact relationship between perturbations and gravity extends at the moment. Indeed, showing that the perturbations to the ef\/fective spacetime structure induced by perturbations of the reference state (e.g., a local restriction of Minkowski vacuum or the cosmological thermal state) correspond to gravitational ef\/fects is perhaps the most important and exciting prospect of our current research.

\section{Conclusion}\label{sec:conclusions}

\subsection{Summary of results}
In Section \ref{sec:framework} we formulated a spacetime-free algebraic framework for describing experimental arrangements with quantum systems. The key motivation for the construction was the observation that the quantum state should be able to alter the causal relations of observables in order to incorporate gravitational phenomena into quantum (f\/ield) theory. Accordingly, the starting point for the formulation was taken to be the \emph{free observable algebra} obtained as the free product $*$-algebra of the abelian von Neumann algebras associated to individual measurements. The other component required to def\/ine an experimental arrangement was identif\/ied as the \emph{reference state} describing the statistical background to measurements. We then explained how the reference state may impose causal relations and, in some cases, also time-evolution on the observables through the GNS representation of the free observable algebra and its modular structure. In particular, the physical observable algebra embodying the dynamical relations was obtained as (the norm completion of) the image of the GNS representation of the free observable algebra induced by the reference state. We also def\/ined the concepts of covariance and symmetry in the framework. Finally, we considered the description of perturbations to the reference state and, in particular, def\/ined the algebra of operators that may be used to induce perturbations. We observed that perturbations of the reference state may causally decouple observables, which was argued to be suggestive of gravitational ef\/fects.

In Section \ref{sec:spacetime} we considered some methods to extract ef\/fective topological and geometric structure from the dynamics of perturbations induced by the reference state. We suggested basing the operational def\/inition of locality on the algebraic and dynamical properties of subalgebras of observables. In particular, local subalgebras were def\/ined as those that are most weakly coupled with their causal complements. Moreover, we argued operational metric information to be obtainable from the readings of quantum clock systems, which we def\/ined. Unfortunately, our discussion had to remain mostly at the level of hypotheses, although these hypotheses seem rather compelling and physically well-motivated to us. In any case, we would like to emphasize that there must undoubtedly exist \emph{some} way to recover topological and geometric information about spacetime from the dynamics of a quantum (f\/ield) system, since it is through the dyna\-mical behavior of matter systems that we determine the structure of spacetime in practice. In the present work, we have proposed only a couple of methods, but other more appropriate and practical ones may exist.

Undoubtedly, the most signif\/icant aspect of the present work is the formulation of the spacetime-free framework for quantum theory that led to the realization of a mechanism for the quantum state to inf\/luence the causal relations and the time-evolution of a quantum system in the spacetime-free framework. The mechanism may allow for the appearance of gravitational phenomena associated to perturbations of the reference state, such as the vacuum, although the exact nature of these ef\/fects remains to be studied more carefully.

\subsection{Challenges to the approach}
Perhaps the most immediate and fundamental challenge for the physical interpretation of the spacetime-free framework for quantum physics, as we have presented it in Section \ref{sec:framework}, is the question of how to determine the reference state. We explained the operational meaning of the reference state in terms of the statistical background to measurements, but in most situations it is clearly not possible to determine the reference state of a system experimentally with suf\/f\/icient accuracy. The problem is most obvious in the case of cosmology, which is partly why we restricted our language to the description of controlled laboratory experiments (the other reason being conceptual clarity). We would like to emphasize that the reference state completely determines the model we have of a system. Therefore, in a sense, the question is like asking for the origin of the gauge group or the equations of motion in f\/ield theory. It may be that the reference state simply has to be a theoretical input to the model in the cases, where it cannot be experimentally deduced. On the other hand, there may exist some universality argument imposing restrictions on the reference state for cosmology arising from, for example, a renormalization property. In particular, it appears reasonable to expect that the `universal' reference state should converge to a pure vacuum state in the limit of the whole universe, whereas in the opposite limit of Planckian systems it should presumably converge to a trace, describing a maximally symmetric inf\/inite temperature state. There may exist a natural choice of a f\/low from one to the other. Maybe it is also worth noting that no physical observer actually has an access to all the observables of the universe, but dif\/ferent observers can access dif\/ferent subsets of the set of all observables. (See, e.g., \cite{Hackl14} for a realization of such a situation in ordinary QFT.) The restriction of the universal quantum state onto the subalgebra of observables accessible to a single observer will introduce non-trivial evolution on the subalgebra, whose properties may be relevant for identifying the appropriate universal state.

Another obvious challenge concerns the recovery of spacetime structure and gravity. As we already emphasized, on physical grounds we expect that there should exist an operational way to recover spacetime structure from the dynamics of a quantum system. Whether the ideas we put forward in this work are up to the task remains to be verif\/ied. Obviously, to determine if the perturbations of the ef\/fective spacetime structure induced by the perturbations of the reference state give rise to gravitational phenomena, one must f\/irst understand how the ef\/fective spacetime structure can be properly studied. However, let us also point out that the universality of the free algebra construction seems to imply that \emph{some} class of perturbations must lead to the right kind of deformation of the ef\/fective spacetime structure.

\subsection{Outlook}
It is clear that a lot of work remains to be done to verify the physical relevance of the spacetime-free approach to quantum physics presented here. In particular, the following two important tasks would bring the goal within reach:
\begin{itemize}\itemsep=0pt
	\item Show explicitly how the ordinary formulation of quantum f\/ield theory on a background spacetime (e.g., f\/lat Minkowski spacetime) can be related to the spacetime-free framework.
	\item Show that spacetime structure can be recovered in the framework to a suf\/f\/icient degree, and f\/ind concrete useful methods to accomplish the recovery.
\end{itemize}
In this work, we have touched upon both of the tasks, but without conclusive results. However, if these challenges can be positively tackled, then we would be able to study the exact relation between the perturbations of the reference state (e.g., the vacuum) and gravitational phe\-no\-mena. This could of\/fer a conceptually coherent and elegant explanation for the emergence of gravity from quantum physics, and open up a vast array of further questions about the physical implications of the framework.

A further important question concerns the emergence of Minkowski spacetime:
\begin{itemize}\itemsep=0pt
	\item Show that the theory leads generically to an approximately f\/lat 4-dimensional ef\/fective local spacetime structure in some appropriate regime.
\end{itemize}
The solution to this problem should probably follow from the statistical properties of quantum states on some free observable algebras, which consist of measurements able to discern spacetime structure. In \cite{Buchholz98} it has been shown that the Poincar\'e symmetry group of f\/lat spacetime is generated in algebraic QFT by the modular involutions associated to subalgebras correspon\-ding to Rindler wedges, but it is not clear at the moment how to translate this result to the spacetime-free framework. Other ideas worth exploring, which have not been mentioned in this work, include the intriguing relation of half-sided modular inclusions to the Lorentz group~\cite{Borchers00}, which might explain the emergence of local spacetime symmetries in more generic situations. Lorentz symmetry has also recently been derived from quantum informational principles~\cite{Hoehn14}. On the other hand, we have not even touched on the question of spacetime dimensionality. The dimension four has many special properties (see, e.g., \cite{Scorpan}), and we cannot help wondering whether these are important. At the same time, the idea that the 3-dimensionality of space is related to the 3-dimensionality of the Bloch ball is an old one (see, e.g.,~\cite{Mueller13, Weizsaecker06}), and worth exploring further in our framework.

Another interesting topic of future research is the exploration and classif\/ication of physical observable algebras obtained from a f\/inite set of binary measurements (and their limits):
\begin{itemize}\itemsep=0pt
	\item Classify states on the free observable algebras $\star_{i=1}^n \mb{C}^2$ for $n\in\mb{N}$ according to the kind of causal structures and evolution that they impose on the observables.
\end{itemize}
(In Appendix \ref{app:examples} we covered some examples with two binary measurements.) This elementary class of free observable algebras describes experimental arrangements with a f\/inite number of binary `yes/no' questions, which covers a wide range of experimental situations. We could even argue that they cover all the practically realizable situations, since in a realistic measurement we always obtain only a f\/inite amount of information \cite{Weizsaecker06}.

All in all, the spacetime-free approach to quantum theory introduced in the current work appears to us as a promising (if still rather hypothetical) new candidate in the collection of attempts to explain the origin of spacetime and gravity. We kindly invite anyone interested to take part in the future work to further elucidate its viability.

\appendix

\section{Probability formula for a sequence of measurements}\label{app:histories}
In the standard Schr\"odinger picture quantum mechanics the state of a closed quantum system is described by a unit state vector in some Hilbert space of vector states $\mc{H}$. The time-evolution of the state vector is generated by the Hamiltonian operator $H$ as $|\psi(t+s)\rangle = U(s)|\psi(t)\rangle \in \mc{H}$, where $U(s) := e^{isH}$ is the one-parameter group of unitaries implementing the time-evolution, and $|\psi(t)\rangle \in \mc{H}$ represents the state of the system at time $t\in\mb{R}$. The probability amplitude for a transition from a vector state $|\psi\rangle\in\mc{H}$ at time $t$ to another vector state $|\xi\rangle\in\mc{H}$ at time $s$ is expressed in the standard way $\langle\xi| U(s-t) |\psi\rangle \in \mb{C}$.

Let us consider further the case, where after transitioning to $|\xi\rangle\in\mc{H}$ at time $s$ (e.g., through a projective measurement), we again let the system to evolve freely, until transitioning yet again to another vector state $|\phi\rangle\in\mc{H}$ at time $r$. By standard quantum mechanics, the probability amplitude for this process may be expressed as the product of the two transition amplitudes
\begin{gather*}
	\langle \phi| U(r-s)|\xi\rangle \langle\xi| U(s-t) |\psi\rangle \in \mb{C} .
\end{gather*}
In general, we may write as
\begin{gather*}
	\mc{A}(|\psi_i\rangle, t_i) := \prod_{i=1}^n \langle\psi_i | U(t_i - t_{i-1}) |\psi_{i-1}\rangle \in \mb{C}
\end{gather*}
the probability amplitude for a process consisting of an initial vector state $|\psi_0\rangle\in\mc{H}$ at time $t_0\in\mb{R}$ followed by projective measurements onto $|\psi_i\rangle\in\mc{H}$ at times $t_i$, $i=1,\ldots,n$.

The probability for the process is obtained as the squared norm of the probability amplitude
\begin{gather*}
	\mc{P}(|\psi_i\rangle, t_i) := |\mc{A}(|\psi_i\rangle, t_i)|^2 = \left( \prod_{i=1}^n \langle\psi_{i-1} | U(t_{i-1} - t_i) |\psi_i\rangle \right) \left( \prod_{j=1}^n \langle\psi_j | U(t_j - t_{j-1}) |\psi_{j-1}\rangle\right) .
\end{gather*}
Let $P_i := |\psi_i\rangle\langle\psi_i|$ denote the projection onto the subspace of $\mc{H}$ spanned by the unit vector $|\psi_i\rangle\in\mc{H}$, which mediates the associated projective measurement. We then obtain
\begin{gather*}
	\mc{P}(|\psi_i\rangle, t_i) = \langle\psi_0| U(t_0) \left( \overrightarrow{\prod_{i=1}^n} U(t_i)^* P_i U(t_i) \right) \left( \overleftarrow{\prod_{i=1}^n} U(t_i)^* P_i U(t_i) \right) U(t_0)^* |\psi_0\rangle ,
\end{gather*}
where an arrow on top of a product sign determines the direction, in which the product index increases along the factors. Here, each of the factors $P_i' := U(t_i)^* P_i U(t_i) = U(t_i)^* |\psi_i\rangle \langle\psi_i| U(t_i)$ is a projection corresponding to the state vector $U(t_i)^* |\psi_i\rangle$, i.e., the intermediate state $|\psi_i\rangle$ translated in time from the transition time $t_i$ to $t=0$. Likewise, $|\psi_0'\rangle := U(t_0)^* |\psi_0\rangle$ is the initial state vector translated from the initial time $t_0$ to $t=0$. Accordingly, we may further rewrite
\begin{gather*}
	\mc{P}(|\psi_i\rangle, t_i) = \langle\psi_0'| \left( \overrightarrow{\prod_{i=1}^n} P_i' \right) \left( \overleftarrow{\prod_{i=1}^n} P_i' \right) |\psi_0'\rangle .
\end{gather*}
Finally, allowing for an arbitrary (possibly mixed) initial state $\omega_0$, we obtain the expression
\begin{gather}\label{eq:processprob}
	\omega_0\big(\big(P_1' P_2' \cdots P_n'\big) \big(P_n' P_{n-1}'\cdots P_1'\big)\big)
\end{gather}
for the probability of a process consisting of projective measurements.\footnote{This standard formula for the probability of a history of events is also the starting point of the consistent histories approach to quantum mechanics (see, e.g.,~\cite{Halliwell94}).} Notice that if two projective measurements corresponding to the projections $P_1$, $P_2$ are performed at two times~$t_1$,~$t_2$, which are related through the time-evolution $P_2 = U(t_2-t_1) P_1 U(t_2-t_1)^*$, then $P_1' = P_2'$ in~(\ref{eq:processprob}), since they correspond to the same operator, when translated to $t=0$:
\begin{gather*}
	P_2' = U(t_2)^*P_2 U(t_2) = U(-t_1) P_1 U(-t_1)^* = U(t_1)^* P_1 U(t_1) = P_1' .
\end{gather*}
In other words, in equation (\ref{eq:processprob}) operators at dif\/ferent times are identif\/ied through the time-evolution. Operators at dif\/ferent times can also always be compared through the identity map if the time-evolution is an isomorphism, but the `same' measurement (as identif\/ied through the identity map) at two dif\/ferent times corresponds (in general) to two dif\/ferent projections in~(\ref{eq:processprob}).

Importantly, in equation (\ref{eq:processprob}) the time parameter does not appear explicitly, but the time-evolution is already accounted for in the def\/initions of the initial state and the projective measurements as being translated to a common instant of time $t=0$. Ultimately, the time-evolution is hidden in the algebraic relations between the projections that are induced by the time-translations. Moreover, we did not impose any ordering of the measurement times $t_i$, so the ordering of the projections in (\ref{eq:processprob}) refers to the order in which the experimenter records the measurement outcomes, and not (necessarily) the temporal order of the measurement events themselves. Therefore, equation (\ref{eq:processprob}) appears promising for generalization to the case, where no global time parameter or evolution is assumed. Indeed, in the main text we will see how a~state can be used to impose algebraic relations between spectral projections, and thus possibly give rise to a~notion causal ordering and evolution.

\section{Positive energy perturbations and gravity}\label{app:positiveenergy}
In this Appendix, we consider a def\/inition of mass for static perturbations, and study if their ef\/fect on the structure of the observable algebra resembles in any sense that of gravity.

We assume in this Appendix that the extended reference state $\tilde\Omega$ is faithful on the physical observable algebra $\mf{A}_\Omega$, and thus induces the modular f\/low $\sigma^\Omega \colon \mb{R} \rightarrow \mt{Aut}(\mf{A}_\Omega)$ as described in Section~\ref{subsec:modular}. If the perturbed reference state $\tilde\Omega'$ is faithful on the perturbed physical observable algebra $\mf{A}_{\Omega'}$, it likewise induces the perturbed modular f\/low $\sigma^{\Omega'}\colon \mb{R} \rightarrow \mt{Aut}(\mf{A}_{\Omega'})$ on $\mf{A}_{\Omega'}$. In this case we may write $\mf{A}_{\Omega'} = P_{\Omega'}\mf{A}_\Omega P_{\Omega'} \subset \mf{A}_\Omega$, where $P_{\Omega'} \in \mf{A}_\Omega'$ is the central projection such that $\{A \in \mf{A}_\Omega \colon \tilde\Omega'(A^*A) = 0\} = \mf{A}_\Omega (\1 - P_{\Omega'})$ (i.e., the support projection of $\tilde\Omega'$). The perturbed modular f\/low $\sigma^{\Omega'}$ is related to the original unperturbed one $\sigma^\Omega$ by a strongly continuous one-parameter family of partial isometries $t \in \mb{R} \mapsto \delta^{\Omega'\Omega}_t \in \mf{A}_\Omega$ called the \emph{Connes cocycle derivative} of $\tilde\Omega'$ with respect to $\tilde\Omega$, which satisf\/ies
\begin{gather*}
	\sigma^{\Omega'}_t(a) = \delta^{\Omega'\Omega}_t \sigma^\Omega_t(a) \big(\delta^{\Omega'\Omega}_t\big)^* \qquad \text{and} \qquad \delta^{\Omega'\Omega}_{s+t} = \delta^{\Omega'\Omega}_s \sigma^\Omega_s\big(\delta^{\Omega'\Omega}_t\big)
\end{gather*}
for all $t,s\in\mb{R}$ and $a\in \mf{A}_{\Omega'} \subset \mf{A}_\Omega$ \cite{Takesaki2}. The cocycle derivative always belongs to the physical observable algebra $\mf{A}_\Omega$, unlike the modular operators, and is therefore (approximately) measurable according to the operational interpretation of our formalism. Physically speaking, the cocycle derivative ref\/lects the dif\/ference in equilibrium dynamics of the two states. In the following, we would like to suggest that (the generator of) the cocycle derivative provides a~measure of energy dif\/ference, or mass, for a perturbed state that is stationary under the thermal (modular) dynamics. However, f\/irst we need to review a few facts about the structure of the GNS Hilbert space.

The norm completion of the set of vectors $\{ a J_\Omega a J_\Omega |\1_\mf{F}\rangle_\Omega \in \mc{H}_\Omega \colon a \in \mf{A}_\Omega \}$ is called the \emph{natural positive cone} $\mc{P}_\Omega \subset \mc{H}_\Omega$, and it has several important algebraic characteristics \cite{Summers06, Takesaki2}. In particular, for any state $\tilde\Omega'$ on $\mf{A}_\Omega$ there exists a unique unit vector $|\psi_{\Omega'}\rangle_\Omega \in \mc{P}_\Omega$ such that $\tilde\Omega'(a) = \langle \psi_{\Omega'}| a | \psi_{\Omega'}\rangle_\Omega$ for all $a\in \mf{A}_\Omega$. Accordingly, the norm completion of the subspace
\begin{gather*}
	\big\{ \pi_\Omega(a) |\psi_{\Omega'}\rangle_\Omega \in \mc{H}_\Omega \colon a \in \mf{F} \big\} \subset \mc{H}_\Omega
\end{gather*}
is isomorphic to the GNS Hilbert space $\mc{H}_{\Omega'}$ induced by the state $\tilde\Omega'$, since $|\psi_{\Omega'}\rangle_\Omega \equiv |\1_\mf{F}\rangle_{\Omega'} \in \mc{H}_{\Omega'}$ is a cyclic vector in $\mc{H}_{\Omega'}$ by def\/inition. Therefore, we may consider the GNS Hilbert space $\mc{H}_{\Omega'}$ of the perturbed state $\Omega'$ as a Hilbert subspace of the reference GNS Hilbert space $\mc{H}_\Omega$ that is invariant under the action of $\pi_\Omega(\mf{F})$. In order to def\/ine the mass operator, we note that the cocycle derivative $\delta^{\Omega'\Omega}_t\colon \mc{H}_\Omega \rightarrow \mc{H}_\Omega$ restricts to an isometry $u^{\Omega'\Omega}_t \colon \Delta_\Omega^{it}\mc{H}_{\Omega'} \rightarrow \mc{H}_{\Omega'}$, since $\ker(\delta^{\Omega'\Omega}_t)^\bot = \Delta_\Omega^{it}\mc{H}_{\Omega'}$.

In the special case that the perturbed state $\tilde\Omega'$ is stationary\footnote{We restrict to consider the stationary case mainly for technical reason. It may be possible to extend the def\/inition of the mass operator to a more general situation, but we have not explored this possibility so far.} under the thermal dynamics given by the modular f\/low $\sigma^\Omega$ induced by the reference state $\tilde\Omega$ (i.e., $\tilde\Omega'(\sigma^\Omega_t(a)) = \tilde\Omega'(a)$ for all $a\in \mf{A}_\Omega$ and $t\in\mb{R}$), we have that $\Delta_\Omega^{it} |\psi_{\Omega'}\rangle_\Omega = |\psi_{\Omega'}\rangle_\Omega$ and thus $\Delta_\Omega^{it}\mc{H}_{\Omega'} = \mc{H}_{\Omega'}$ for all $t\in\mb{R}$. Accordingly, $u^{\Omega'\Omega}_t\colon \mc{H}_{\Omega'} \rightarrow \mc{H}_{\Omega'}$ are unitary, and form a one-parameter group (i.e., \mbox{$u^{\Omega'\Omega}_t u^{\Omega'\Omega}_s = u^{\Omega'\Omega}_{t+s}$ for all $t,s\in\mb{R}$}). We may then consider the generator $h^{\Omega'\Omega}$ of the unitaries $u^{\Omega'\Omega}_t \equiv \exp(-ith^{\Omega'\Omega})$, which is a densely def\/ined self-adjoint operator af\/f\/iliated with \mbox{$\mf{A}_{\Omega'}\subset B(\mc{H}_{\Omega'})$}. We suggest to interpret $h^{\Omega'\Omega}$ as measuring the physical energy/mass content associated with the perturbation. In particular, $\Omega'$ is considered to represent a positive mass perturbation with respect to the background def\/ined by the reference state $\Omega$, if $h^{\Omega'\Omega}$ is a positive operator. Notice that stationary perturbations are mapped to stationary perturbations by the symmetries of the reference state $\tilde\Omega$. If $\tilde\Omega'$ is faithful on the unperturbed observable algebra~$\mf{A}_\Omega$, then $h^{\Omega'\Omega}$ simply gives the dif\/ference of the generators of the two dynamics (i.e., Hamiltonians), with respect to which the two states are in equilibrium. However, we must allow $\tilde\Omega'$ not to be faithful in order for its GNS representation to have a non-trivial kernel, and thus alter the algebraic structure of the physical observable algebra.

We conjecture the following physically interesting feature of the above mathematical structure: If the mass operator $h^{\Omega'\Omega}$ is positive, then $\ker \pi_\Omega \subset \ker \pi_{\Omega'}$ is a proper inclusion. Below, we will provide a sketch for a proof in the f\/inite-dimensional case, but a rigorous proof is left for future work. If this property holds true, it implies physically that perturbations with positive mass always turn some new measurements jointly measurable. This seems analogous of the property of gravity that a positive mass perturbation focuses lightcones, and therefore always makes some previously timelike separated (not jointly measurable) local subsystems spacelike separated (jointly measurable).

Let the perturbed state $\Omega'$ be stationary under the reference modular f\/low $\sigma^\Omega$. Here we provide a sketch of a proof in the f\/inite-dimensional case for the fact that $\ker \pi_\Omega \subset \ker \pi_{\Omega'}$ must be a proper inclusion if the mass operator $h^{\Omega'\Omega}$ is strictly positive.

Let $\rho_\Omega, \rho_{\Omega'} \in B(\mc{H}_\Omega)$ be the density operators corresponding to the reference state $\Omega$ and its perturbation $\Omega'$ on $\mf{A}_\Omega$, respectively. Since $\Omega'$ is stationary under $\sigma^\Omega$, which is represented by the adjoint action of $\rho_\Omega^{it}$, the density operators $\rho_\Omega \in B(\mc{H}_\Omega)$ and $\rho_{\Omega'} \in B(\mc{H}_\Omega)$ must have a~common basis of eigenvectors in $\mc{H}_\Omega$. In fact, we have $u^{\Omega'\Omega}_t = \rho_{\Omega'}^{it}\rho_\Omega^{-it}$. For the genera\-tor~$h^{\Omega'\Omega}$ of~$u^{\Omega'\Omega}_t$ to be a strictly positive operator, each non-vanishing eigenvalue of~$\rho_{\Omega'}$ must be larger than the corresponding eigenvalue of $\rho_\Omega$, because the eigenvalues of~$h^{\Omega'\Omega}$ are logarithms of ratios of the non-zero eigenvalues of the two density operators. On the other hand, both~$\Omega$ and~$\Omega'$ are normalized, i.e., the eigenvalues of both $\rho_\Omega$ and $\rho_{\Omega'}$ sum up to $1$. Therefore, some of the eigenvalues of $\rho_{\Omega'}$ must vanish, i.e., the support of $\Omega'$ is non-trivial. Accordingly, $\ker \pi_\Omega \subset \ker \pi_{\Omega'}$ is a proper inclusion.

\section{Experiments with two binary measurements}\label{app:examples}
Let us consider the simplest non-trivial example of the framework we have presented in Section \ref{sec:framework}: an experimental arrangement, where we have access to only two measurements, which can both take only two values each. The measurements are thus represented by two abelian von Neumann algebras $\mf{W}_i \cong \mb{C}^2$, $i=1,2$, each linearly spanned by two spectral projections $P^{(i)}_k \in \mf{W}_i$, $k=1,2$, which satisfy $P^{(i)}_k P^{(i)}_l = \delta_{kl} P^{(i)}_k$ and $\sum_k P^{(i)}_k = \1$. The free product $*$-algebra $\mf{F} \cong \mb{C}^2 \star \mb{C}^2$ is linearly spanned by f\/inite sequences of spectral projections
\begin{gather*}
	P^{(i_1)}_{k_1} P^{(i_2)}_{k_2} \cdots P^{(i_n)}_{k_n} \in \mf{F} ,
\end{gather*}
where $i_m \neq i_{m+1}$ for all $m=1,\ldots,n-1$, $k_m=1,2$ for all $m=1,\ldots,n$, and $n\in\mb{N}$, which represent sequences of measurement outcomes. Actually, due to the relation $P^{(i)}_2 = \1 - P^{(i)}_1$ for $i=1,2$, only the sequences of the form
\begin{gather*}
	P^{(i_1)}_1 P^{(i_2)}_1 \cdots P^{(i_n)}_1 \in \mf{F}
\end{gather*}
are linearly independent in $\mf{F}$.

So far we are considering abstract elements of an abstract algebra. The reference state gives the physical meaning to the formalism by assigning the observational probabilities to dif\/ferent measurement outcome sequences. Therefore, it is interesting to consider dif\/ferent kind of states on $\mf{F}$ and the kinds of physical observable algebras that they may induce.

\subsection*{States on $\boldsymbol{\mb{C}^2\otimes\mb{C}^2}$}

As the simplest example, a state $\omega$ on the tensor product $\mb{C}^2\otimes\mb{C}^2$ can be lifted onto $\mf{F}$ via the pull-back of the algebra homomorphism $\phi\colon \mf{F} \cong \mb{C}^2\star\mb{C}^2 \rightarrow \mb{C}^2 \otimes \mb{C}^2$ def\/ined by
\begin{gather*}
	 \1_\mf{F} \stackrel{\phi}{\mapsto} \1^{(1)} \otimes \1^{(2)} \equiv \1_{\mb{C}^2 \otimes \mb{C}^2} ,\\
	 P^{(1)}_1 \stackrel{\phi}{\mapsto} P^{(1)}_1 \otimes \1^{(2)} ,\\
	 P^{(2)}_1 \stackrel{\phi}{\mapsto} \1^{(1)} \otimes P^{(2)}_1 , \qquad \text{and}\\
	 P^{(i_1)}_1 \cdots P^{(i_n)}_1 \stackrel{\phi}{\mapsto} P^{(1)}_1 \otimes P^{(2)}_1
\end{gather*}
for any sequence such that $i_m\neq i_{m+1}$ for all $m=1,\ldots,n-1$ and $n>1$. Then, the pull-back state $\Omega := \omega \circ \phi\colon \mf{F} \rightarrow \mb{C}$ satisf\/ies
\begin{gather}
	 \Omega(\1_\mf{F}) = 1 ,\nonumber\\
	 \Omega\big(P^{(1)}_1\big) \in [0,1] ,\nonumber\\
	 \Omega\big(P^{(2)}_1\big) \in [0,1] ,\nonumber\\
	 \Omega\big(P^{(i_1)}_1 \cdots P^{(i_n)}_1\big) = \Omega\big(P^{(1)}_1 P^{(2)}_1\big) \in [0,1] \label{eq:tensorstate}
\end{gather}
for any sequence such that $i_m\neq i_{m+1}$ and $n>1$, so the expectation values on the basis elements of $\mf{F}$ form a very simple pattern. In particular, all the expectation values are uniquely determined in terms of $\Omega\big(P^{(1)}_1\big)$, $\Omega\big(P^{(2)}_1\big)$ and $\Omega\big(P^{(1)}_1 P^{(2)}_1\big)$.

Let us assume that the state $\omega$ is faithful on $\mb{C}^2\otimes\mb{C}^2$. However, $\Omega$ is clearly \emph{not} faithful on the free algebra $\mb{C}^2\star\mb{C}^2$. Indeed, $\big[P^{(1)}_k,P^{(2)}_l\big] \in \mc{N}_\Omega := \{a\in \mf{F}\colon \Omega(a^*a) = 0\}$ for any $k,l=1,2$ as is easily verif\/ied by using the expectation values of $\Omega$ above, and therefore we have $\big|P^{(2)}_lP^{(1)}_k \big\rangle_\Omega \sim \big|P^{(1)}_kP^{(2)}_l \big\rangle_\Omega$ in the GNS Hilbert space $\mc{H}_\Omega$. It is also straightforward to verify, e.g., that
\begin{gather*}
	\big| P^{(1)}_{k_1} P^{(2)}_{l_1} P^{(1)}_{k_2} P^{(2)}_{l_2} \cdots P^{(1)}_{k_n} P^{(2)}_{l_n} \big\rangle_\Omega \sim \big| P^{(1)}_{k_1} P^{(2)}_{l_1} \big\rangle_\Omega
\end{gather*}
in $\mc{H}_\Omega$ if $k_m = k_{m+1}$ and $l_m = l_{m+1}$ $\forall\, m$, and $\sim 0$ otherwise. Similar equivalence relations apply between other sequences of length $> 2$ and sequences of length $2$ (or the null vector). Accordingly, we obtain the GNS Hilbert space $\mc{H}_\Omega \cong \mb{C}^4$ with the orthonormal basis of vectors
\begin{gather*}
	e_{kl} := \frac{1}{\sqrt{\Omega\big(P^{(1)}_k P^{(2)}_l\big)}} \big|P^{(1)}_kP^{(2)}_l \big\rangle_\Omega \in \mc{H}_\Omega ,\qquad k,l=1,2 .
\end{gather*}
The cyclic vector is given by
\begin{gather*}
	|\1_\mf{F}\rangle_\Omega = \sum_{k,l} \big|P^{(1)}_kP^{(2)}_l\big\rangle_\Omega = \sum_{k,l} \sqrt{\Omega\big(P^{(1)}_k P^{(2)}_l\big)} e_{kl} \in \mc{H}_\Omega
\end{gather*}
in terms of the orthonormal basis. We have likewise $\pi_\Omega(\mb{C}^2\star\mb{C}^2) \cong \mb{C}^2\otimes\mb{C}^2$ for the GNS representation. Thus, the inf\/inite-dimensional non-abelian free observable algebra $\mb{C}^2\star\mb{C}^2$ is reduced to the f\/inite-dimensional abelian physical observable algebra $\mb{C}^2\otimes\mb{C}^2$ by the statistics~(\ref{eq:tensorstate}) of the reference state~$\Omega$. The tensor product structure of the physical observable algebra indicates the operational independence of the two measurements.

The modular structure induced by the reference state $\Omega$ is rather trivial, because the physical observable algebra is abelian: We have $\Delta_\Omega = \1$ and $J_\Omega = C$, the complex conjugation operator. The perturbation algebra is given by $B(\mb{C}^4) \cong B(\mb{C}^2) \otimes B(\mb{C}^2)$. Perturbations of the reference state $\Omega \mapsto \Omega'$ such that $\Omega'(a) = \sum_n \Omega(b_n^* a b_n)$, $b_n \in B(\mb{C}^4)$, for all $a\in\mf{A}_\Omega \cong \mb{C}^2\otimes\mb{C}^2$ may alter the sta\-tis\-tics~(\ref{eq:tensorstate}) by changing the expectation values $\Omega\big(P^{(1)}_1\big)$, $\Omega\big(P^{(2)}_1\big)$ and $\Omega\big(P^{(1)}_1 P^{(2)}_1\big)$, but they cannot introduce non-trivial commutation relations between the observables. Therefore, also all physical perturbations have trivial modular structure. We must proceed to more involved examples in order to recover non-trivial thermal dynamics.

\subsection*{States on $\boldsymbol{\mb{M}_2(\mb{C})}$}
We may consider a homomorphism $\phi\colon \mb{C}^2\star\mb{C}^2 \rightarrow \mb{M}_2(\mb{C})$, the algebra of 2-by-2 complex-valued matrices, identifying the projections $P^{(i)}_k \in \mb{C}^2\star\mb{C}^2$ with two arbitrary pairs of complementary projections given by $P^{(i)}_1 = \big|u^{(i)}\big\rangle \big\langle u^{(i)}\big| \in \mb{M}_2(\mb{C})$ and $P^{(i)}_2 = \1_2 - P^{(i)}_1$, where $\big|u^{(i)}\big\rangle := \big(u^{(i)}_1,u^{(i)}_2\big) \in\mb{C}^2$ is an arbitrary vector of unit norm. We then simply f\/ind
\begin{gather*}
	 \big(P^{(1)}_1 P^{(2)}_1\big)^{n} = \big| \big\langle u^{(1)} | u^{(2)} \big\rangle \big|^{2(n-1)} P^{(1)}_1 P^{(2)}_1 , \\
	 \big(P^{(1)}_1 P^{(2)}_1\big)^{n} P^{(1)}_1 = \big| \big\langle u^{(1)} | u^{(2)} \big\rangle \big|^{2n} P^{(1)}_1 , \\
	 P^{(2)}_1 \big(P^{(1)}_1 P^{(2)}_1\big)^{n} = \big| \big\langle u^{(1)} | u^{(2)} \big\rangle \big|^{2n} P^{(2)}_1 .
\end{gather*}
We then consider a state $\omega$ on $\mb{M}_2(\mb{C})$ given by $\omega(a) = \tr(\rho a)$, where $\rho \in \mb{M}_2(\mb{C})$ is a positive matrix with unit trace, and again set $\Omega = \omega\circ\phi\colon \mb{C}^2\star\mb{C}^2 \rightarrow \mb{C}$ as our reference state. By the above, we f\/ind
\begin{gather*}
	 \Omega\big(\big(P^{(1)}_1 P^{(2)}_1\big)^{n}\big) = \lambda^{n-1} \Omega\big(P^{(1)}_1 P^{(2)}_1\big) \qquad \forall\, n=1,2,\ldots ,\nonumber\\
	 \Omega\big(\big(P^{(1)}_1 P^{(2)}_1\big)^{n} P^{(1)}_1 \big) = \lambda^{n} \Omega\big(P^{(1)}_1\big) \qquad \forall\, n=0,1,\ldots ,\nonumber\\
	 \Omega\big(P^{(2)}_1 \big(P^{(1)}_1 P^{(2)}_1\big)^{n}\big) = \lambda^{n} \Omega\big(P^{(2)}_1\big) \qquad \forall\, n=0,1,\ldots ,
\end{gather*}
where
\begin{gather}
	\lambda = \big| \big\langle u^{(1)} | u^{(2)} \big\rangle \big|^2 \in [0,1] ,\nonumber\\
	\Omega\big(P^{(i)}_1\big) = \big\langle u^{(i)} | \rho | u^{(i)} \big\rangle \in [0,1] ,\qquad i = 1,2 ,\nonumber\\
	\Omega\big(P^{(1)}_1 P^{(2)}_1\big) = \big\langle u^{(1)} | u^{(2)} \big\rangle \big\langle u^{(2)} | \rho | u^{(1)} \big\rangle \in \mb{C} . \label{eq:M2staterel}
\end{gather}
Thus, all the expectation values are determined by $\Omega\big(P^{(i)}_1\big)$, $\Omega\big(P^{(1)}_1 P^{(2)}_1\big)$ and the parameter $\lambda$, which are related through~(\ref{eq:M2staterel}).

It is interesting to study the possible equivalence relations that we may have in this case for the GNS Hilbert space $\mc{H}_\Omega \equiv \overline{\mf{F}/\mc{N}_\Omega}$. For example, we f\/ind by a direct calculation that
\begin{gather*}
	\Omega\Big( \big| \big(P^{(1)}_1 P^{(2)}_1\big)^n - P^{(1)}_1 P^{(2)}_1 \big|^2 \Big) = \lambda \big(\lambda^{n-1} - 1\big)^2 \Omega\big(P^{(2)}_1\big) ,
\end{gather*}
where we introduced the notation $|a|^2 := a^*a$. We have that $\big(P^{(1)}_1 P^{(2)}_1\big)^n - P^{(1)}_1 P^{(2)}_1 \in \mc{N}_\Omega$, and therefore $\big|\big(P^{(1)}_1 P^{(2)}_1\big)^n\big\rangle_\Omega \sim \big|P^{(1)}_1 P^{(2)}_1\big\rangle_\Omega$ in $\mc{H}_\Omega$ if the above expression vanishes. But this happens only if either (i) $\lambda = 0$, which implies $P^{(1)}_1 = \1 - P^{(2)}_1 \equiv P^{(2)}_2$, or (ii)~$\lambda = 1$, which implies $P^{(1)}_1 = P^{(2)}_1$. Both of the cases reduce the observable algebra into the abelian~$\mb{C}^2$, and therefore are rather trivial. On the other hand, we may compute for a constant $\alpha\in\mb{C}$
\begin{gather*}
	\Omega\Big( \big| P^{(1)}_1 P^{(2)}_1 P^{(1)}_1 - \alpha P^{(1)}_1 \big|^2 \Big) = |\lambda - \alpha|^2 \Omega\big(P^{(1)}_1\big) ,
\end{gather*}
so we see that $\big|P^{(1)}_1 P^{(2)}_1 P^{(1)}_1 \big\rangle_\Omega \sim \lambda \big|P^{(1)}_1\big\rangle_\Omega$ in $\mc{H}_\Omega$, and similarly $\big|P^{(2)}_1 P^{(1)}_1 P^{(2)}_1 \big\rangle_\Omega \sim \lambda \big|P^{(2)}_1\big\rangle_\Omega$. Actually, we further have that all sequences of projections are equivalent to linear combinations of sequences of length $\leq 2$, since $\mc{N}_\Omega$ is a left-ideal. Moreover, sequences of length $<2$ may be expressed as linear combinations of sequences of length~$2$ (e.g., $P^{(1)}_1 = P^{(1)}_1P^{(2)}_1 + P^{(1)}_1P^{(2)}_2$), and therefore the elements $\big|P^{(i)}_kP^{(j)}_l\big\rangle_\Omega \in \mc{H}_\Omega$ span the GNS Hilbert space. The cyclic vector is given in terms of these vectors as
\begin{gather*}
	|\1_\mf{F}\rangle_\Omega = \sum_{k,l} \big|P^{(i)}_kP^{(j)}_l\big\rangle_\Omega
\end{gather*}
for any $i,j=1,2$, $i\neq j$. However, not all of them are linearly independent, because we have the relations
\begin{gather*}
	\sum_l \big|P^{(i)}_kP^{(j)}_l\big\rangle_\Omega = \big|P^{(i)}_k\big\rangle_\Omega = \sum_l \big|P^{(j)}_l P^{(i)}_k\big\rangle_\Omega
\end{gather*}
for all $i\neq j$. Let us assume that $\lambda \in (0,1)$ and $\Omega\big(P^{(i)}_1\big), \Omega\big(P^{(1)}_1 P^{(2)}_1\big) \in (0,1)$ for $i=1,2$, i.e., $\omega$~is faithful on $\mb{M}_2(\mb{C})$. Then, starting for example with the orthonormal vectors $\big|P^{(1)}_k\big\rangle_\Omega \in \mc{H}_\Omega$, $k=1,2$, and applying the Gram--Schmidt method to f\/ind the rest of the basis will leave us with four orthonormal basis vectors, and accordingly $\mc{H}_\Omega \cong \mb{C}^4$, as for the usual GNS construction with respect to a faithful state on $\mb{M}_2(\mb{C})$. The modular operator $S_\Omega$ def\/ined by $S_\Omega |a\rangle_\Omega = |a^*\rangle_\Omega$ gives rise to the modular structure. We have that, given an orthonormal basis, $S_\Omega$ may be represented by a non-singular matrix in $\mb{M}_4(\mb{C})$ composed with a complex conjugation, which takes care of the anti-linearity of the map $|a\rangle_\Omega \stackrel{S_\Omega}{\mapsto} |a^*\rangle_\Omega$. The polar decomposition of this matrix will give us representations of the modular operators $\Delta_\Omega$ and $J_\Omega$ acting on $\mc{H}_\Omega \cong \mb{C}^4$, which are generically non-trivial in this case.

\subsection*{Free product states}
The free product of states plays an important role in the study of free product algebras and free independence in non-commutative probability theory \cite{Voiculescu92}. In our case, a free product state $\Omega$ is def\/ined by the following centralizing property:
\begin{gather}\label{eq:centralized}
	\Omega\big(\big(P^{(i_1)}_1 - \Omega\big(P^{(i_1)}_1\big)\big) \big(P^{(i_2)}_1 - \Omega\big(P^{(i_2)}_1\big)\big) \cdots \big(P^{(i_n)}_1 - \Omega\big(P^{(i_n)}_1\big)\big)\big) = 0
\end{gather}
for all sequences such that $i_m \neq i_{m+1}$ for all $m=1,\ldots, n-1$ and $n\in\mb{N}$. Let us call $n$ the order of the expectation value $\Omega\big(P^{(i_1)}_1 P^{(i_2)}_1 \cdots P^{(i_n)}_1\big)$. Since we may write~(\ref{eq:centralized}) as
\begin{gather*}
	\Omega\big(P^{(i_1)}_1 P^{(i_2)}_1 \cdots P^{(i_n)}_1\big) + \sum (\mt{products of expectation values of order} < n) = 0 ,
\end{gather*}
it is actually possible to solve for the expectation values recursively to all orders in terms of the f\/irst order expectation values $\Omega\big(P^{(i)}_1\big)$, $i=1,2$. However, the explicit expression in terms of~$\Omega\big(P^{(i)}_1\big)$ grow rapidly in complexity as a function of the order~\cite{Voiculescu92}. For the few lowest orders we f\/ind
\begin{gather*}
	\Omega\big(P^{(1)}_1 P^{(2)}_1\big) = \Omega\big(P^{(1)}_1\big) \Omega\big(P^{(2)}_1\big) , \\
	\Omega\big(P^{(1)}_1 P^{(2)}_1 P^{(1)}_1\big) = \Omega\big(P^{(2)}_1 P^{(1)}_1 P^{(2)}_1\big) = \Omega\big(P^{(1)}_1\big) \Omega\big(P^{(2)}_1\big) , \\
	\Omega\big(P^{(1)}_1 P^{(2)}_1 P^{(1)}_1 P^{(2)}_1\big) = \Omega\big(P^{(1)}_1\big) \Omega\big(P^{(2)}_1\big) \big[ \Omega\big(P^{(1)}_1\big) + \Omega\big(P^{(2)}_1\big) - \Omega\big(P^{(1)}_1\big) \Omega\big(P^{(2)}_1\big) \big] , \\
	\Omega\big(P^{(1)}_1 P^{(2)}_1 P^{(1)}_1 P^{(2)}_1 P^{(1)}_1\big) = \Omega\big(P^{(1)}_1\big) \Omega\big(P^{(2)}_1\big) \big[ \Omega\big(P^{(1)}_1\big) + \Omega\big(P^{(2)}_1\big) - \Omega\big(P^{(1)}_1\big)^2 \\
\hphantom{\Omega\big(P^{(1)}_1 P^{(2)}_1 P^{(1)}_1 P^{(2)}_1 P^{(1)}_1\big) =}{}
 - 4 \Omega\big(P^{(1)}_1\big) \Omega\big(P^{(2)}_1\big) + 4 \Omega\big(P^{(1)}_1\big)^2 \Omega\big(P^{(2)}_1\big) \big].
\end{gather*}
Despite their complexity, the free product states are known to satisfy a few important properties\footnote{For the proofs of these and other interesting properties, see, e.g.,~\cite{Avitzour82,Dykema98,Dykema98b} and the references therein.}:
\begin{itemize}\itemsep=0pt
	\item If the restrictions of a free product state $\Omega$ onto the free product factors are faithful, $\Omega$ is faithful on the free product. Therefore, $\Omega$ gives rise to a faithful GNS representation of $\mb{C}^2\star\mb{C}^2$, and the measurements are `freely independent'.
	\item If the restrictions of a free product state $\Omega$ onto the free product factors are tracial, $\Omega$ is tracial on the free product. Since all states on $\mb{C}^2$ (or any other abelian algebra) are tracial, a free product state $\Omega$ provides a normalized trace on $\mb{C}^2\star\mb{C}^2$. Consequently, the physical observable algebra is of type II$_1$ according to the classif\/ication of von Neumann algebras.
	\item Since a free product state $\Omega$ on $\mb{C}^2\star\mb{C}^2$ is a trace, its modular automorphism group is trivial. However, non-free perturbations of $\Omega$ may induce non-trivial dynamics.
\end{itemize}
Interestingly, tensor product states and free product states are opposite extremes in the con\-ti\-nuum of states on $\mb{C}^2\star\mb{C}^2$ in the following sense: The f\/irst ones lead to a fully commutative algebra, while the second ones lead to a maximally non-commutative algebra of physical obser\-vables.

\section{Identif\/ication of local subsystems in spin lattices}\label{app:spinlattices}
The dynamics of a closed quantum system described by the observable algebra $\mf{A}$ is given by a one-parameter family of automorphisms $\sigma\colon \mb{R} \rightarrow \mt{Aut}(\mf{A})$. The magnitude of correlations between two subalgebras $\mf{B}_{1},\mf{B}_2\in\mf{A}$ at times $0$ and $t$, respectively, can be quantif\/ied by
\begin{gather*}
	\mc{C}_t(\mf{B}_1,\mf{B}_2) := \sup_{\substack{b_1\in\mf{B}_1\\b_2\in\mf{B}_2}} \frac{\| [\sigma_t(b_1), b_2] \|}{2 \|b_1\| \|b_2\|} \in [0,1] .
\end{gather*}
The Lieb--Robinson theorem \cite{Eisert2014, Lieb72} sets a bound on the magnitude of correlations for a certain class of quantum systems, whose dynamics are generated by time-independent Hamiltonians, which are approximately local with respect to some background geometry, such as a lattice. A~simple form of the bound reads
\begin{gather*}
	\mc{C}_t(\mf{B}_1,\mf{B}_2) \leq C e^{-\mu(d_{12} - v|t|)} ,
\end{gather*}
where \looseness=1 $d_{12}$ is the spatial distance between the two subsystems represented by $\mf{B}_{1,2}$, and \mbox{$C, \mu, v {\in}\mb{R}_+$} are f\/inite constants, whose values depend on the dynamics. In particular, $v$ is called the Lieb--Robinson velocity, which is the (approximate) maximum speed for information transfer in the system, analogous to the speed of light in relativistic systems. Due to the Lieb--Robinson bound, the correlations between subsystems decay exponentially outside the ef\/fective `lightcone'.

The correlations between a subsystem represented by $\mf{B}$ and its environment $\mf{B}'$ in the total system are measured by $\mc{C}_t(\mf{B},\mf{B}')$. Thus, the growth of correlations induced by the dynamics may be quantif\/ied by
\begin{gather*}
	\mc{D}(\mf{B}) := \left. \frac{\dd}{\dd t} \right|_{t=0} \mc{C}_t(\mf{B},\mf{B}') .
\end{gather*}
Let $\mf{B}\subset \mf{A}$ be any subfactor of the physical observable algebra. We then compare the values of~$\mc{D}(\alpha(\mf{B}))$ for dif\/ferent $\alpha \in \mt{Aut}(\mf{A})$. We conjecture that a subalgebra $\alpha(\mf{B}) \subset \mf{A}$, for which this quantity acquires a minimum should correspond to a local spherical subsystem.

Taking advantage of the Lieb--Robinson theorem, it is not dif\/f\/icult to see why $\mc{D}(\mf{B})$ is minimized by local subsystems for at least some simple non-relativistic quantum systems with local Hamiltonians, although a more rigorous proof could certainly be deviced: A~global auto\-morphism $\alpha\in\mt{Aut}(\mf{A})$ rendering a local observable algebra less local can alternatively be represented by a corresponding global transformation of the Hamiltonian, $H \mapsto \alpha^{-1}(H)$. As the dynamics becomes less local, the transformation increases the Lieb--Robinson group velocity, to which~$\mc{D}(\mf{B})$ is roughly proportional.

We have explored these intuitions via computer simulations of simple f\/inite-dimensional quantum systems. For the simulations we have used instead of $\mc{C}_t(\mf{B}_1,\mf{B}_2)$ the quantity
\begin{gather*}
	\mc{E}_t(\mf{B}_1, \mf{B}_2) := \frac{1}{2 d_1^2 d_2^2} \sum_{i,j,k,l} \big\| \big[\sigma_t(e_{ij}^1), e_{kl}^2\big] \big\|_{\rm HS} \in [0,1]
\end{gather*}
to \looseness=1 estimate the magnitude of commutators between two subalgebras, where $d_{1,2} = \dim(\mf{B}_{1,2})$, $\|b\|_{\rm HS} \equiv \sqrt{\tr(b^*b)}$ denotes the Hilbert--Schmidt norm in the fundamental representation of $\mf{A}$, and the summation runs over the bases of matrices of the form $(e_{ij}^{1,2})_{mn} = \delta_{im}\delta_{jn}$. The normaliza\-tion removes any direct dependence on the dimensions of the two subalgebras. $\mc{E}_t(\mf{B}_1, \mf{B}_2)$ is more readily computable than the supremum of the commutator norm over all operators, while behaves qualitatively very similarly, since they both measure the magnitude of commutators.

\begin{figure}[t]
\centering
\includegraphics[width=75mm]{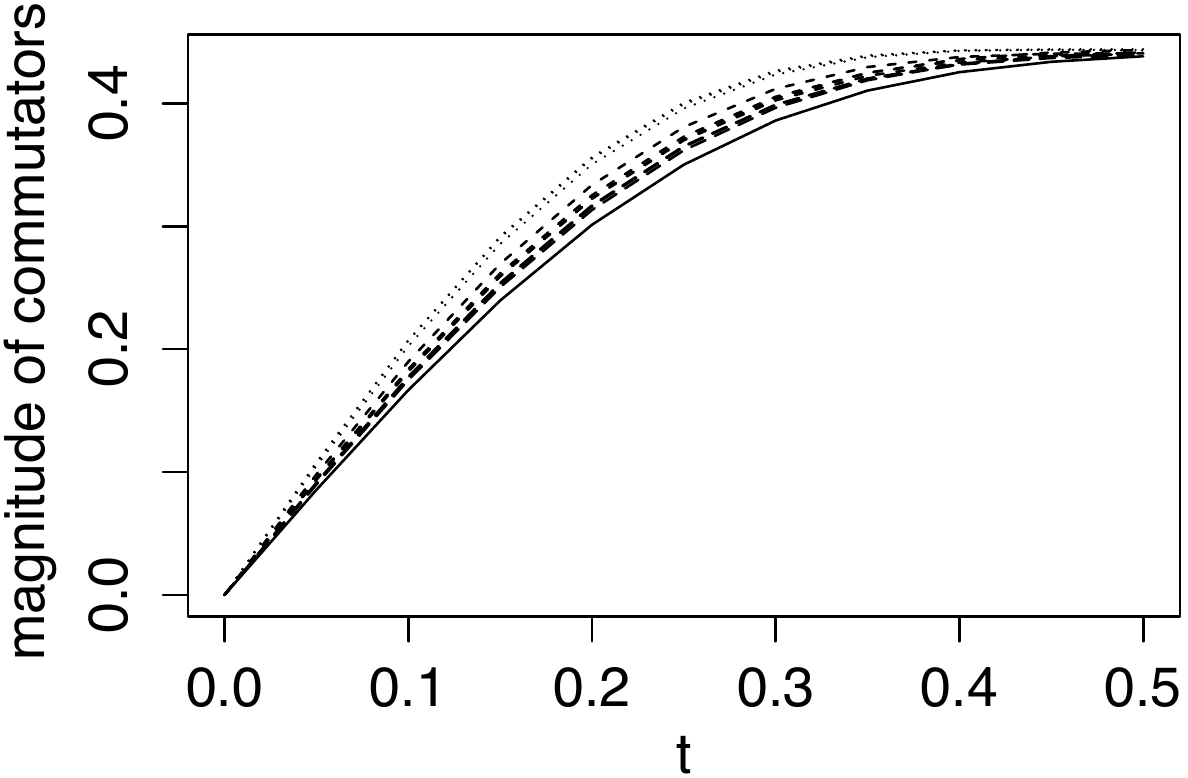}\hfill\includegraphics[width=70mm]{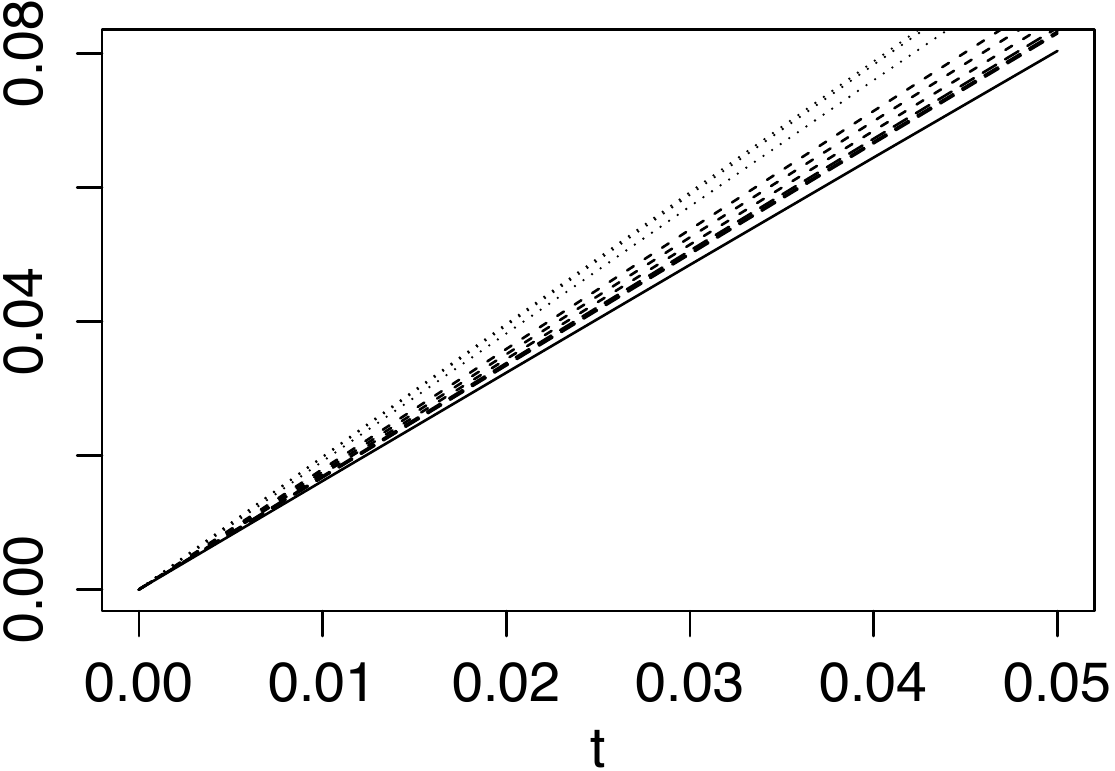}
\caption{Evolution of $\mc{E}_t(\mf{B}, \mf{B}')$ for increasingly non-local subsystems.}\label{fig:4chain_spinenv}
\end{figure}

\looseness=1 In Fig.~\ref{fig:4chain_spinenv} we illustrate in two dif\/ferent time-scales two typical examples of our simulation results for the time-evolution of $\mc{E}_t(\mf{B}, \mf{B}')$ for a subalgebra $\mf{B}$ with its commutant~$\mf{B}'$.\footnote{All of our computations were performed in the R environment. The code used in our `computational experiments' can be found at~\url{http://github.com/Oct8poid/ReconGeom}.} The plots in Fig.~\ref{fig:4chain_spinenv} concern a ring of four $\frac{1}{2}$-spins. The Hamiltonian was chosen to be a sum of local pairwise terms of the form $c_{ij}\sigma_i^{k}\sigma_j^{k+1\ (\mt{mod } 4)}$, where $\sigma_i^k$ is the $i$'th Pauli matrix ($i=1,2,3$) of the $k$'th spin system ($k=1,\ldots,4$), and the coef\/f\/icients $c_{ij}$ were chosen at random from a uniform probability distribution in the interval $[-1,1]$. The solid lines correspond to a strictly local subalgebra, namely, any one of the spins. The dashed lines correspond to non-local subalgebras obtained from the local one by a random global unitary transform. The unitary transforms were generated by Hermitian matrices, whose elements were chosen at random from a uniform distribution in the interval $[0,d]$, where the upper bound $d$ took values $0.1$, $0.2$ and $0.4$ indicated by the density of gaps in the lines in increasing order, thus varying the magnitude of the non-local perturbation. We see clearly in our simulations that the growth of commutators is invariably the slowest for local subalgebras, and becomes faster as we strengthen the non-local perturbation. Dif\/ferent local dynamics and dif\/ferent background topologies reproduce very consistent outcomes.

\subsection*{Acknowledgments}
First and foremost, I would like to thank Carlos AdS/CFT Guedes for introducing me to the algebraic approach to quantum f\/ield theory, and for many inspiring discussions in the very early stages of the work. Likewise, I would like to thank Paolo Bertozzini and Roberto Conti for supporting and inf\/luencing the development of the ideas in this manuscript. The comments and suggestions by Philipp H\"ohn and Sebastian Steinhaus were very helpful during the preparation of this manuscript. Many thanks also to Miklos L{\aa}ngvik, Ted Jacobson and Klaus Fredenhagen for instructive discussions. Furthermore, I would like to express my gratitude to the anonymous referees for their feedback, which helped to improve this manuscript signif\/icantly. Finally, I am indebted to Maria Kalimeri for her assistance with the R language, among other things. This work has been generously funded by the Finnish social security system.

\pdfbookmark[1]{References}{ref}
\LastPageEnding

\end{document}